\documentclass{ws-ijmpa}
\usepackage[super,compress]{cite}
\usepackage[caption=false]{subfig}
\usepackage{graphicx}
\usepackage[breaklinks]{hyperref}
\hypersetup{colorlinks,urlcolor=black,citecolor=black,linkcolor=black,filecolor=black}
\begin{document}
\markboth{Malace, Gaskell, Higinbotham, Clo{\"e}t}{The Challenge of the EMC Effect}

%
\catchline{}{}{}{}{}
%

\title{The Challenge of the EMC Effect: existing data and future directions}

\author{Simona Malace, David Gaskell, Douglas W. Higinbotham}
\address{Thomas Jefferson National Accelerator Facility\\
Newport News, VA, 23601, USA}

\author{Ian~C.~Clo\"et}
\address{Argonne National Laboratory, Argonne, Illinois 60439, USA}

\maketitle


\begin{abstract}
Since the discovery that the ratio of inclusive charged lepton (per-nucleon) cross sections
from a nucleus $A$ to the deuteron is not unity -- even in deep inelastic scattering kinematics -- a 
great deal of experimental and theoretical effort has gone into understanding the phenomenon.   
The EMC effect, as it is now known, shows that even in the most extreme kinematic conditions the 
effects of the nucleon being bound in a nucleus can not be ignored. 
In this paper we collect the most precise data available for various nuclear to deuteron 
ratios, as well as provide a commentary on the current status of the theoretical understanding of 
this thirty year old effect.
\end{abstract}

\ccode{PACS numbers:
~24.85.+p, 
~13.60.Hb, 
~25.30.Fj, 
~25.30.Mr  
}

\tableofcontents

\section{Introduction}

Scientific understanding sometimes gets sidetracked when a simple model works
better than one might \textit{a priori} expect.  This was the case with the
nuclear independent particle shell models which did an excellent job predicting
the excitation levels of many nuclei as well as predicting the functional form
of many cross sections.  This success has led many groups, even to this day, to
think of nucleons in the nucleus as independent particles in a mean field
potential.  
Most recently, this paradigm affected the neutrino community, where comparison of 
recent results to Fermi gas models of the nucleus led to an initial 
surprise~\cite{AguilarArevalo:2008rc,AguilarArevalo:2010wv}. These initial results then motivated subsequent 
work using sophisticated descriptions of nucleons in the nucleus for the analysis of neutrino scattering
results~\cite{Ankowski:2012ei,Morfin:2012kn,Golan:2013jtj}.

In this review, we return to one of the most investigated cases of protons and
neutrons in the nucleus not behaving as free nucleons; the EMC (European Muon
Collaboration) effect.  In the original experiment 120-280\,GeV muons were
scattered from an iron target and then compared to deuterium data~\cite{Aubert:1983xm}.  Plots of the
extracted structure function ratios as a function of the Bjorken scaling variable for the proton ($x_p$) surprised
many~\cite{Aubert:1983xm}.  In these deep inelastic kinematics, the per-nucleon
cross section ratio was not unity. Even in kinematics where the underlying
degree of freedom should be quarks and gluons, effects of the nucleon being in
the nucleus were still playing a role.

Over the years many EMC type experiments have been performed and a number of
experimental conclusions were drawn: most notably, that the shape of the effect
was universal, that the functional form was relatively $Q^2$ independent, and
that the effect slowly and simply increased with the $A$ of the
nucleus~\cite{Gomez:1993ri,Adams:1995is,Amaudruz:1995tq,Arneodo:1996qe},
exhibiting an $A$ dependence that was consistent both with $\log{A}$ and average
nuclear density.  Except for Drell-Yan type experiments, it seemed that the
experimental part of the EMC effect story was complete; leaving theorists to
sort out the EMC puzzle with data that was easily described qualitatively, but
more difficult to achieve detailed agreement over the full $x$ range.

In 2009 the story changed due to new high precision measurements of the EMC
effect on light nuclei~\cite{Seely:2009gt}.  This data clearly disagreed with a
simple logarithmic $A$ or average nuclear density dependence. Instead, the EMC
effect suddenly appeared to depend on the local density or the cluster structure
of the nucleus as demonstrated by variational Monte Carlo calculations of
$^9$Be~\cite{Pieper:2002ne}.  This one new bit of information has reinvigorated the experimental and
theoretical efforts to pin down the underlying cause of the EMC effect.

In this review, we have collected the data on the EMC effect and present it in a
largely phenomenological way.  We also present the new observation of a
correlation between $x>1$ nuclear data and deep inelastic scattering data.
Finally, we will present a summary of modern theoretical
interpretations of these phenomena, as well as describe the future measurements
needed to further elucidate the origins of the EMC effect.

\section{Review of Data}

In this section we review the world measurements of the ratio of nuclear to
deuterium cross sections from inclusive charged lepton scattering.  The goal is
to assess to what degree the $x$-dependence and precision of current data
constrain the nuclear medium modifications of the structure inside
nuclei for individual nuclear targets.

\subsection{Formalism \label{sec:formalism}}
Charged lepton--nucleon deep inelastic scattering (DIS) is a very powerful tool
for studying the structure of the nucleon. The electromagnetic interaction
governs the coupling of the charged lepton to the nucleon via exchange of
virtual photons.  In Quantum Electrodynamics, the charged lepton--virtual photon
vertex is exactly calculable. The coupling of the virtual photon to the nucleus
is described by the hadronic tensor which can depend on several structure
functions~\cite{Jaffe:1988up,Cloet:2006bq}. The experimental mapping of these functions in a wide kinematic range
and their description in a consistent theoretical model are aimed at
understanding the makeup of the nucleus as seen by the electromagnetic probe.
In the one-photon exchange approximation, the cross section for charged
lepton--nucleon scattering from the proton can be written as:
\begin{equation}
\frac{d^{2}\sigma}{d\Omega dE^{'}} = \frac{\alpha^{2}}{Q^{4}} \frac{E^{'}}{E} L_{\mu \nu} W^{\mu \nu} = \frac{4 \alpha^{2} (E^{'})^{2}}{Q^{4}} \cos^{2} \frac{\theta}{2} \times \left (\frac{F_{2}(x,Q^{2})}{\nu} + \frac{2 F_{1}(x,Q^{2})}{M} \tan^{2} \frac{\theta}{2} \right),
\label{shit}
\end{equation}
where $\alpha$ is the fine structure constant, $Q^{2} =
4EE^{'}\sin^{2}(\theta/2)$ is the four-momentum transfer squared, $E$ and
$E^{'}$ are the initial and scattered charged lepton energies, $L_{\mu \nu}$ and
$W^{\mu \nu}$ are the leptonic and hadronic tensors, $\nu = E - E^{'}$ is the
energy transfer, $x = \frac{Q^{2}}{2M\nu}$ is the Bjorken scaling variable,
$\theta$ is the detected lepton scattering angle while $M$ is the nucleon mass.

In the quark--parton model the structure functions $F_{1}$ and $F_{2}$ are 
expressed in terms of the quark and anti-quark distribution functions as:

\begin{equation}
F_{2}(x) = 2 x F_{1} = x \sum_{q} e_{q}^{2} (q(x) + \bar{q}(x)),
\end{equation}
where $e_q$ are the quark charges and $q(x)$ gives the probability to strike a quark of 
flavor $q$ inside the nucleon carrying a lightcone momentum fraction $x$ of the nucleon momentum. 
Beyond the quark-parton model the 
$Q^{2}$ dependence of the structure functions arises from perturbative Quantum 
Chromodynamics radiative effects as well as from non-perturbative $1/Q^{2}$ power 
corrections. 


In the context of experimental observables, 
the total cross section can be expressed in terms of the absorption cross section
of purely longitudinal ($\sigma_{L}$) and transverse ($\sigma_{T}$) photons:
\begin{equation}
\frac{d^{2}\sigma}{d\Omega dE^{'}} = \Gamma \left(\sigma_{T}(x,Q^{2}) + \epsilon\, \sigma_{L}(x,Q^{2}) \right) = \Gamma \sigma_{T}\,(1 + \epsilon R),
\end{equation}
where $R = \sigma_{L}/\sigma_{T}$. The flux of transverse virtual photons is given by
\begin{equation}
\Gamma = \frac{\alpha}{2 \pi^{2} Q^{2}} \frac{E^{'}}{E} \frac{K}{1 - \epsilon}, 
\end{equation}
where $K = \nu (1-x)$ and the ratio of the longitudinal to transverse virtual photon 
polarizations can be expressed as
\begin{equation}
\epsilon = \Big[ 1 + 2 \left(1 + \frac{\nu^{2}}{Q^{2}} \right) \tan^{2} \frac{\theta}{2} \Big]^{-1}.
\end{equation}

The structure functions can then be written in terms of the experimental 
cross sections as:

\begin{equation}
F_{1}(x,Q^{2}) = \frac{K}{4 \pi^{2} \alpha} M \sigma_{T}(x,Q^{2}) and
\end{equation}

\begin{equation}
F_{2}(x,Q^{2}) = \frac{K}{4 \pi^{2} \alpha} \frac{\nu}{(1 + \nu^{2}/Q^{2})} [\sigma_{T}(x,Q^{2}) + \sigma_{L}(x,Q^{2})].
\end{equation}
In order to extract the $F_{1}$ and $F_{2}$ structure functions from cross 
section measurements, a separation of the longitudinal and transverse 
contributions to the total cross section is required. This $L/T$ separation 
is typically done 
experimentally by employing the Rosenbluth technique~\cite{Rosenbluth:1950yq} 
which involves acquiring 
measurements at two but preferably more $\epsilon$ values at fixed 
$x$ and $Q^{2}$ and then performing a linear fit to the reduced cross section, 
$\frac{d^{2}\sigma}{d\Omega dE^{'}} \frac{1}{\Gamma}$, to extract $\sigma_{T}$ 
and $\sigma_{L}$. This is a specialized type of measurement where a variety 
of experimental configuration changes (beam energies, scattering angles, 
scattered lepton energies) 
are required in a short time interval. Most model-independent $L/T$ separations 
have been performed on a hydrogen and at times deuterium target. There are few 
measurements of $R = \sigma_{L}/\sigma_{T}$ in nuclei and the question of 
whether $R$ would be modified in the nuclear medium has not been 
conclusively answered. Typically the measured cross section ratios 
$(\sigma_{A}/A)/(\sigma_{D}/2)$ and $F_{2}^{A}/F_{2}^{D}$ are assumed to be 
identical but this is true only in the limit of $\epsilon = 1$ or  
$R_{A} - R_{D} = 0$ as illustrated by:
\begin{equation}
\frac{\sigma_A}{\sigma_D} = \frac{F_{2}^{A}(x,Q^2)}{F_{2}^{D}(x,Q^2)} \frac{1 + R_{D}}{1 + R_{A}} 
 \frac{1 + \epsilon R_{A}}{1 + \epsilon R_{D}} \approx \frac{F_{2}^{A}(x,Q^2)}{F_{2}^{D}(x,Q^2)} 
\left[1 - \frac{\Delta R(1-\epsilon)}{(1 + R_D)(1 + \epsilon R_D)}  \right].
\label{eq:cs_tof2}
\end{equation}

When nuclei are probed in the DIS regime, the per--nucleon 
ratio of the structure function $F_{2}$ between an isoscalar nucleus and 
deuterium is viewed as a measure of nuclear modifications of quark distributions 
in nuclei. For a non--isoscalar nucleus an additional correction is applied, the 
so called isoscalar correction, which accounts for the difference in the DIS cross 
sections between protons and neutrons. This correction can be written as: 
\begin{equation}
f_{ISO}^{A} = \frac{\frac{1}{2} \left(1+\frac{F_{2}^{n}}{F_{2}^{p}}\right)}
                 {\frac{1}{A}\left[Z+(A-Z)\frac{F_{2}^{n}}{F_{2}^{p}}\right]},
\label{eq:iso_cor}
\end{equation}
and depends on the neutron to proton $F_{2}$ structure function ratio. 
As there are no free neutron targets, 
the $F_{2}^{n}$ extraction and thus the isoscalar correction is model dependent. A recent 
study~\cite{daniel2007} utilized several parametrizations for $F_{2}^{n}/F_{2}^{p}$ to assess 
the sensitivity of the isoscalar correction to the $n/p$ prescription used. This study employed 
$F_{2}^{n}/F_{2}^{p}$ extracted from SLAC \cite{Bodek:1979rx} and NMC \cite{Arneodo:1995cq,Arneodo:1996qe} proton and 
deuterium DIS measurements as well as $F_{2}^{n}/F_{2}^{p}$ constructed from parton distribution 
functions from CTEQ \cite{Lai:2007dq}. SLAC applied Fermi motion corrections to the deuterium data and 
extracted an unsmeared $F_{2}^{n}$ while NMC made no corrections for nuclear effects. Both 
collaborations neglected possible binding effects in deuterium. CTEQ also neglected the Fermi motion 
of nucleons inside the nucleus. It was found that the isoscalar correction for $^3$He and $^{197}$Au 
varies by at most 3.5$\%$ when different parametrizations for $n/p$ are used. The correction itself 
can reach 10--15$\%$ at $x$ of 0.8 for a gold target.

\subsection{EMC Effect Measurements}

Qualitatively, the dependence on Bjorken $x$ of the per--nucleon ratio of
inclusive charged lepton cross sections from a nucleus $A$ to the deuteron is
well known. For small $x$ values, in the {\it shadowing region}, $x < 0.05 -
0.1$, the ratio is suppressed, the suppression increasing with increasing $A$
and decreasing $x$. For $0.1 < x < 0.3$, the {\it antishadowing region}, the
ratio appears to be enhanced with a hint of some, but no obvious $A$ dependence.
In the interval $0.3 < x < 0.8$, the {\it EMC effect} region, the ratio is
suppressed, its slope in $x$ generally increasing with $A$.  Finally, for $x > 0.8$ the
ratio increases dramatically above unity and this is ascribed to nucleon motion
inside the nucleus (Fermi motion). Various models attempt to describe the
nuclear modification of the experimental ratio $(\sigma_{A}/A)/(\sigma_{D}/2)$
but there is no comprehensive understanding of the entire pattern. Alongside
efforts to construct a satisfactory theoretical model, the experimental
investigation of nuclear targets with lepton probes in the DIS region has continued
over the years and yielded an impressive body of data.  Whether the available
measurements are sufficient to satisfactorily constrain the pattern of nuclear
modifications of nucleon structure needs to be re-evaluated.

\begin{table}[htbp]
\tbl{World measurements of lepton deep inelastic scattering cross section ratios on nuclear targets to deuterium.}
{\begin{tabular}{|c|c|c|c|c|c|c|}
\hline
Target & Collaboration/ &  Ref. & Beam & Energy & Point-to-point & Norm. \\
       & Laboratory   &  &      &  (GeV)      & uncert. (\%) & uncert. (\%)\\
\hline \hline
$^3$He & JLab & \citen{Seely:2009gt} & e & 6 & 1.1 - 2.4  & 1.84 \\
       & HERMES & \citen{hermes_he3_n:2003} & $ e $ & 27 & 1.1 - 2.5 & 1.4 \\
\hline
$^4$He & JLab & \citen{Seely:2009gt} & e & 6 & 1.1 - 2.3  & 1.5 \\
       & SLAC & \citen{Gomez:1993ri} & e & 8 - 24.5 & 1.8 - 12.4 & 2.42 \\
       & NMC  & \citen{Amaudruz:1995tq} & $\mu$ & 200 & 1 - 8.1 & 0.4 \\
\hline 
$^6$Li & NMC & \citen{Arneodo:1995cs} & $\mu$ & 90 & 0.9 - 13.6 & 0.4 \\
\hline
$^9$Be & JLab & \citen{Seely:2009gt} & e & 6 & 1.2 - 2.1 & 1.7 \\
       & SLAC & \citen{Gomez:1993ri} & e & 8 - 24.5 & 0.94 - 10.4  & 1.22  \\
       & NMC  & \citen{Arneodo:1996rv,Amaudruz:1995tq} & $\mu$ & 200 & 1.5 - 8 & 0.45 \\
\hline
$^{12}$C & JLab & \citen{Seely:2009gt} & e & 6 & 1.2 - 2.2 & 1.6 \\
        & SLAC & \citen{Gomez:1993ri} & e & 8 - 24.5 & 1 - 4.5 & 1.22 \\
        & NMC  & \citen{Amaudruz:1995tq} & $\mu$ & 200 & 1 - 7.4 & 0.4 \\
        & EMC  & \citen{Arneodo:1989sy} & $\mu$ & 280 & 5.6 - 9.1 & 7 \\
\hline
$^{14}$N & HERMES & \citen{hermes_he3_n:2003} & $ e $ & 27 & 1 - 3.6 & 1.4 \\
        & BCDMS & \citen{Bari:1985ga} & $\mu$ & 280 & 1.7 - 6.5 & 1.3 \\
\hline
$^{27}$Al & Rochester-SLAC-MIT & \citen{Bodek:1983ec} & e & 4.5 - 20 & 1.3 - 50 & 2.3 \\
       & SLAC & \citen{Gomez:1993ri} & e & 8 - 24.5 & 0.9 - 10.3  & 1.22  \\
       & NMC  & \citen{Arneodo:1996rv,Amaudruz:1995tq} & $\mu$ & 200 & 1.6 - 9.4  & 0.45 \\
\hline
$^{40}$Ca & SLAC & \citen{Gomez:1993ri} & e & 8 - 24.5 & 1.2 - 5.9 & 1.35 \\
         & NMC  & \citen{Amaudruz:1995tq} & $\mu$ & 200 & 0.9 - 7.3 & 0.4 \\
        & EMC  & \citen{Arneodo:1989sy} & $\mu$ & 280 & 5.1 - 10.2 & 7 \\
\hline
$^{56}$Fe & Rochester-SLAC-MIT & \citen{Bodek:1983qn} & e & 4.5 - 20 & 1.7 - 21.9 & 1.1  \\
       & SLAC & \citen{Gomez:1993ri} & e & 8 - 24.5 & 0.9 - 9.7 & 1.4  \\
       & NMC  & \citen{Arneodo:1996rv,Amaudruz:1995tq} & $\mu$ & 200 & 1.5 - 7.8  & 0.45 \\
       & BCDMS  & \citen{Benvenuti:1987az} & $\mu$ & 200 & 1.2 - 4.6  & 1.5 \\
\hline
$^{64}$Cu & EMC & \citen{Ashman:1992kv} & $\mu$ & 100 - 280 & 1.3 - 4 & - \\
\hline
$^{108}$Ag & SLAC & \citen{Gomez:1993ri} & e & 8 - 24.5 & 1.3 - 6.3 & 1.49 \\
\hline
$^{119}$Sn & NMC & \citen{Arneodo:1996rv,Amaudruz:1995tq} & $\mu$ & 200 & 1.1 - 7  & 0.45 \\
          & EMC & \citen{Ashman:1988bf} & $\mu$ & 100 - 280 & 4.4 - 10.3  & 0.9 \\
\hline
$^{197}$Au & SLAC & \citen{Gomez:1993ri} & e & 8 - 24.5 & 1.1 - 12.9 & 2.51 \\
\hline
$^{207}$Pb & NMC & \citen{Arneodo:1996rv,Amaudruz:1995tq} & $\mu$ & 200 & 1.7 - 9.2 & 0.45 \\
\hline
\end{tabular} \label{tab:data} }
\end{table}

We performed an analysis of the world's data 
to extract from measurements the $x$--dependence of the nuclear medium
modifications of the nucleon structure from a variety of nuclear targets. We
utilized the published world data on $(\sigma_{A}/A)/(\sigma_{D}/2)$ as outlined
in Table~\ref{tab:data} folding into our analysis both the point-to-point and the
normalization uncertainties.
The goal was to assess to what degree the kinematic coverage and the precision of current 
data constrain the x-dependence of the nuclear medium modifications of the nucleon 
structure inside nuclei for individual nuclear targets.
Our procedure, which used a Monte Carlo Technique, is described below.

In an initial step for each data set (for a given target) we independently
generate random numbers (a collective of $N$ random numbers for each data set)
that are distributed according to a Gaussian distribution with mean of 1 and standard
deviation equal to the normalization uncertainty. For each individual data set
we then create $N$ pseudo-data collectives by scaling the respective data (the
ratio and the point-to-point uncertainty) by the random numbers generated. For a
given individual $N$ we then combine the pseudo-data sets thus generated from
each experiment and we perform a global fit using a flexible functional
form to account for possible tensions between data from different
experiments. The choice of the fit function is not inspired by physics but
rather it is intended to be flexible enough so that the global fit obtained will
be driven by the quality and quantity of data and not by its functional
form. For nuclear targets where there is only one experimental data set
available we used a $3^{\text{rd}}$ order polynomial while for the rest a $6^{\text{th}}$
order polynomial was used. In the end the global fit that describes the combined
experimental data sets when taking into account the normalization uncertainty is
the average of the $N$ fits to the combined pseudo-data sets and its uncertainty
is the standard deviation of this collective:
\begin{align}
\label{ave}
F(x) &= \frac{\sum^{N}_{i=1} F_{i}(x)}{N},   \\
\label{sd}
\delta F(x) &= \sqrt{ \frac{\sum^{N}_{i=1} (F_{i}(x) - F(x))^2}{N-1}}.
\end{align}
In a last step we also folded the point-to-point uncertainties in the 
fitting procedure by generating $N$ random numbers for each data point in 
each data set that distribute according to a Gaussian with mean of 1 and 
standard deviation equal to the point-to-point uncertainty. The randomization 
driven by the point-to-point uncertainties is applied directly to the 
pseudo-data collectives corresponding to each experiment that has
been previously obtained by randomizing the data according to their 
normalization uncertainty. Then each of the $N$ combined pseudo-data sets 
are fit and a global fit with an uncertainty is extracted 
for each nuclear target according to Eqs.~(\ref{ave}) and (\ref{sd}).

\begin{figure}[htbp]
\centering\includegraphics[width=\columnwidth]{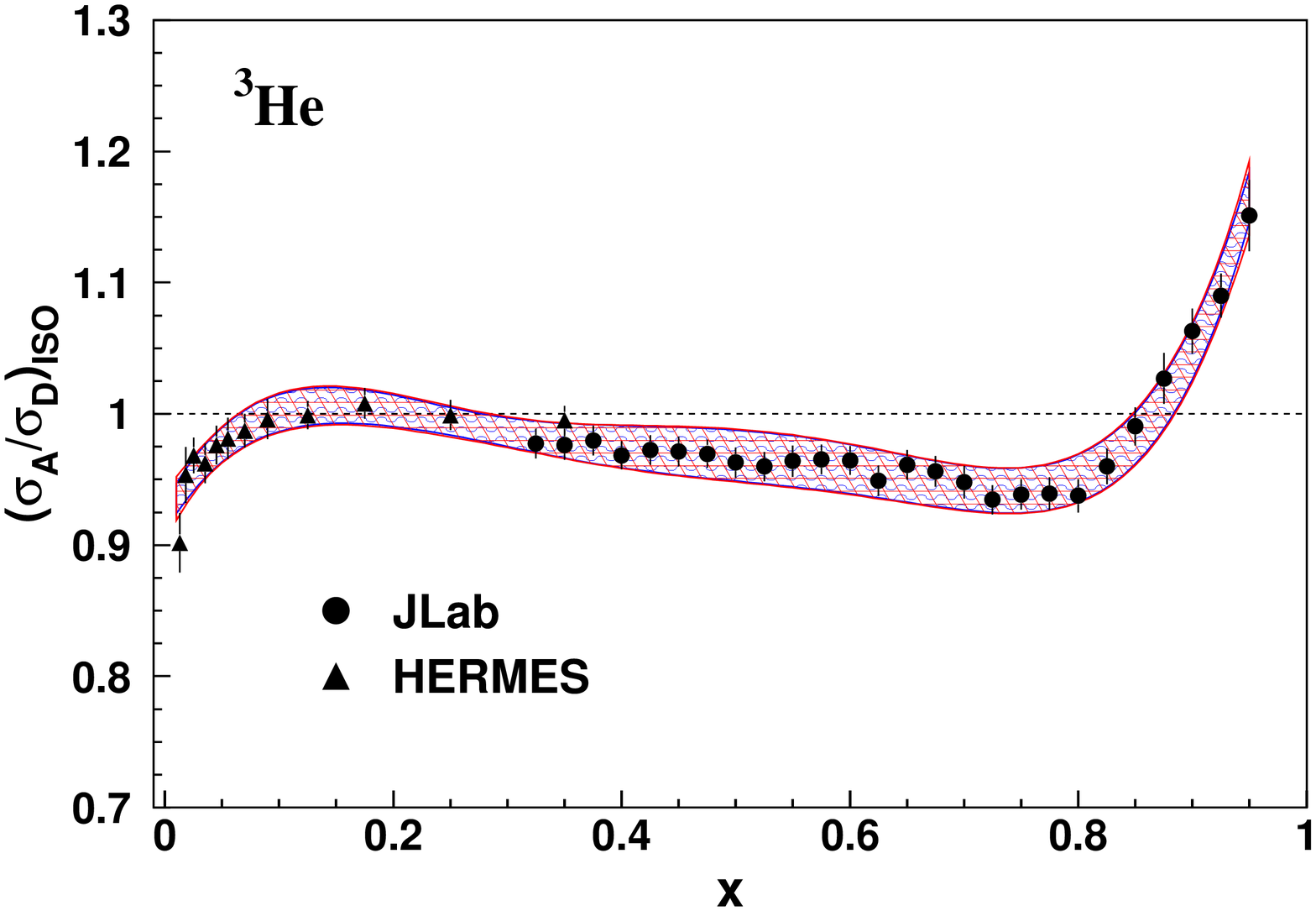}\\
\centering\includegraphics[width=\columnwidth]{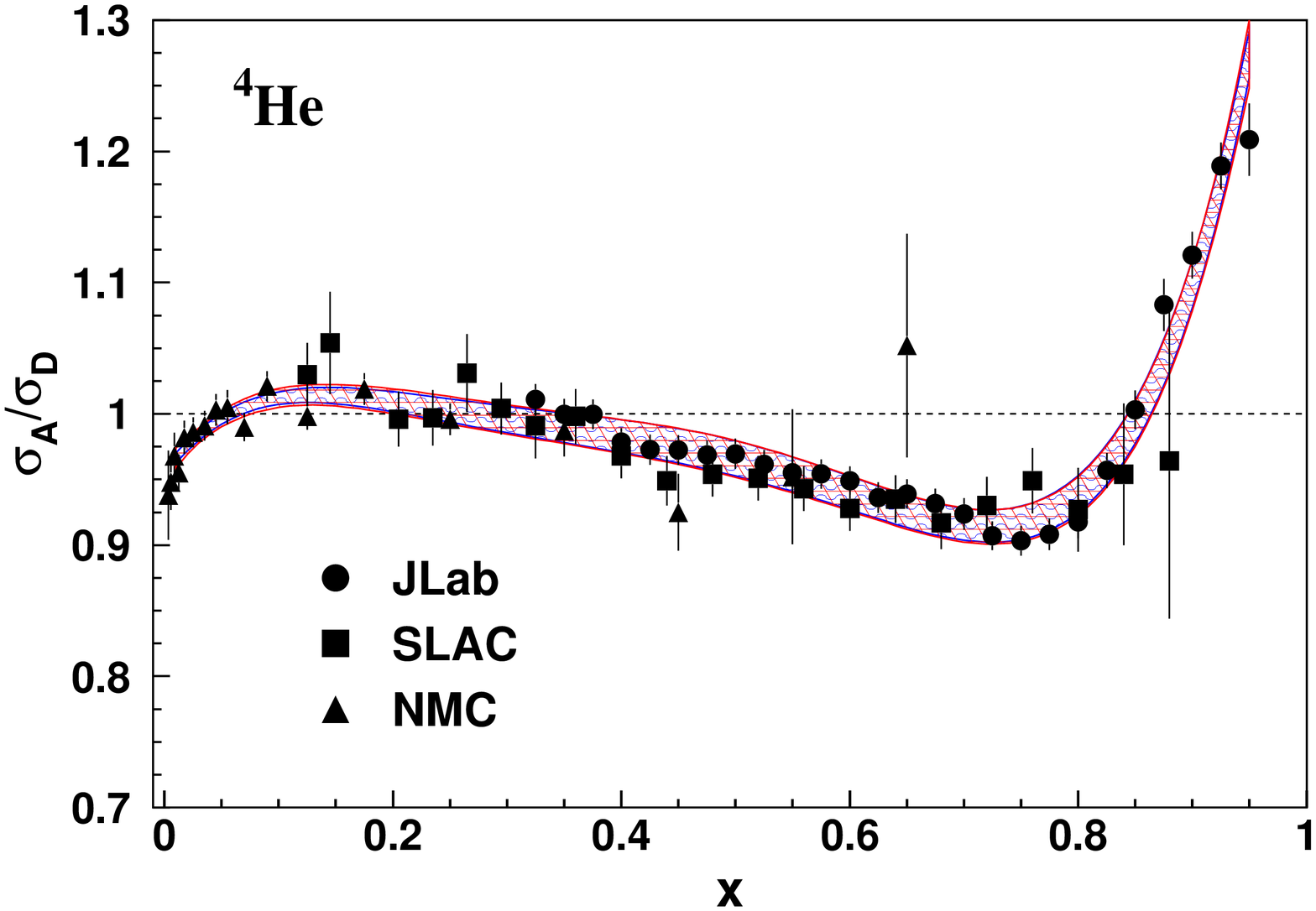} 
\caption{ {\bf Top:} Ratios of the $^3$He to Deuterium DIS cross sections. 
The measurements shown are from JLab~\cite{Seely:2009gt} (circles) 
and HERMES~\cite{hermes_he3_n:2003} (triangles). {\bf Bottom:} Ratios of the 
$^4$He to Deuterium DIS cross sections. The measurements shown are from 
JLab~\cite{Seely:2009gt} (circles), NMC~\cite{Amaudruz:1995tq} (triangles) 
and SLAC~\cite{Gomez:1993ri} (squares). In both panels the error bars 
represent the statistical and point-to-point systematic uncertainties added in 
quadrature. The bands show the fits to the combined data 
sets and their uncertainties when both the normalization and point-to-point 
errors are taken into account (see text for details).}
\label{fig:he3_he4}
\end{figure}

\begin{figure}[htbp]
\centering\includegraphics[width=\columnwidth]{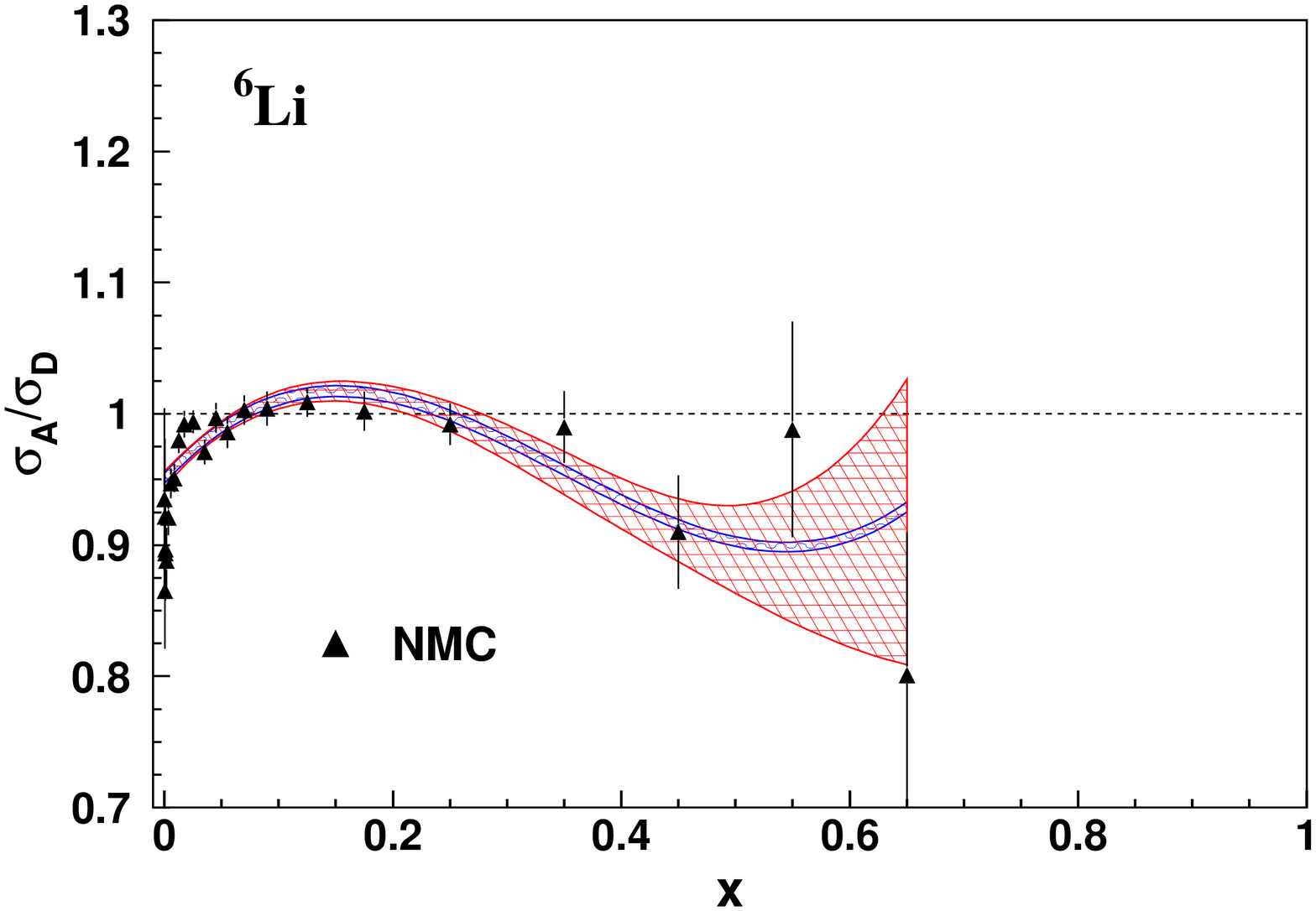}\\
\centering\includegraphics[width=\columnwidth]{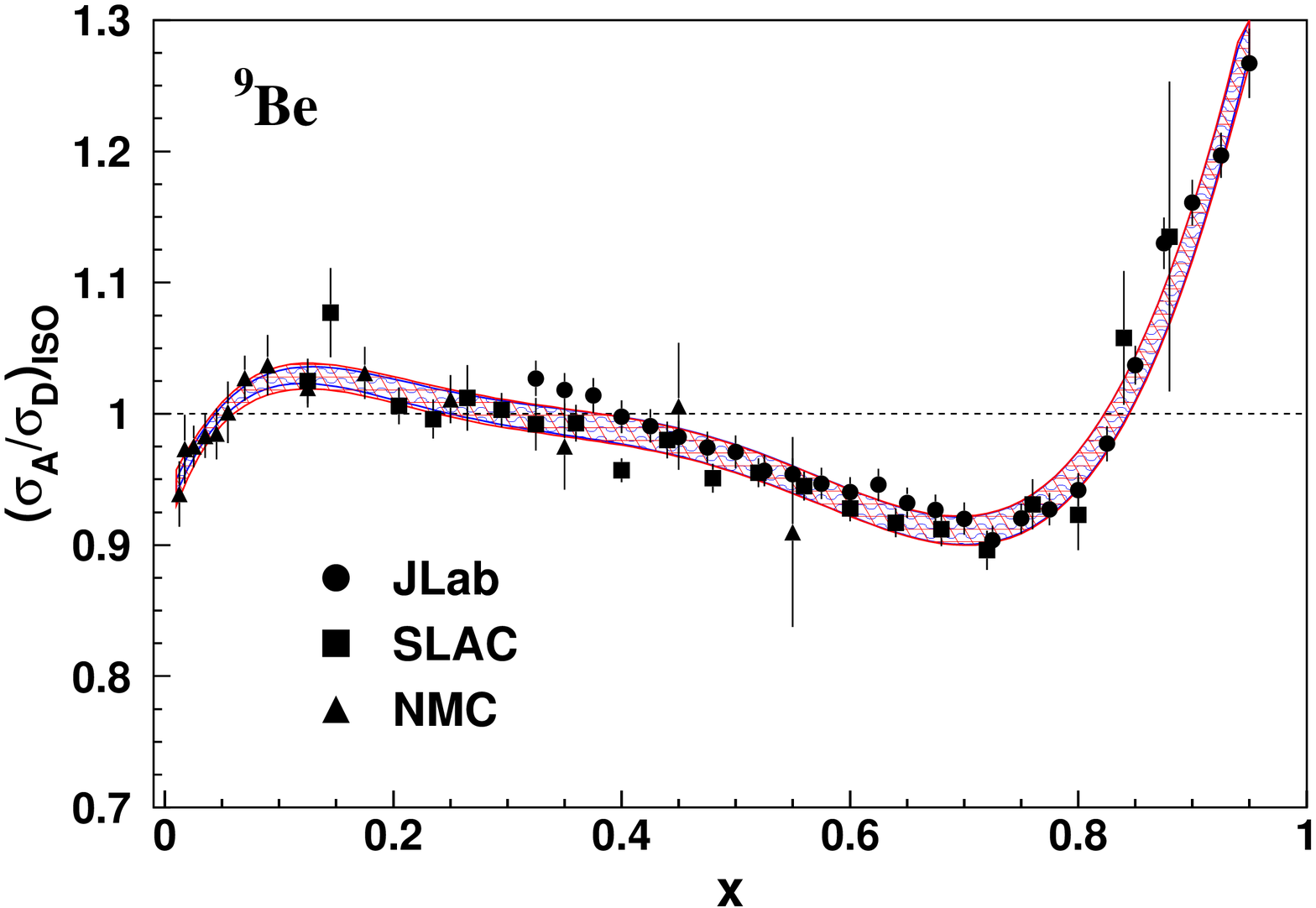} 
\caption{{\bf Top:} Ratios of the $^6$Li to Deuterium DIS cross sections. 
The measurements shown are from NMC~\cite{Arneodo:1995cs} (triangles). 
{\bf Bottom:} Ratios of the 
$^9$Be to Deuterium DIS cross sections. The measurements shown are from 
JLab~\cite{Seely:2009gt} (circles), NMC~\cite{Arneodo:1996rv,Amaudruz:1995tq} 
(triangles) 
and SLAC~\cite{Gomez:1993ri} (squares). In both panels the error bars 
represent the statistical and point-to-point systematic uncertainties added in 
quadrature. The bands show the fits to the combined data 
sets and their uncertainties when both the normalization and point-to-point 
errors are taken into account (see text for details).}
\label{fig:li_be}
\end{figure}

\begin{figure}[htbp]
\centering\includegraphics[width=\columnwidth]{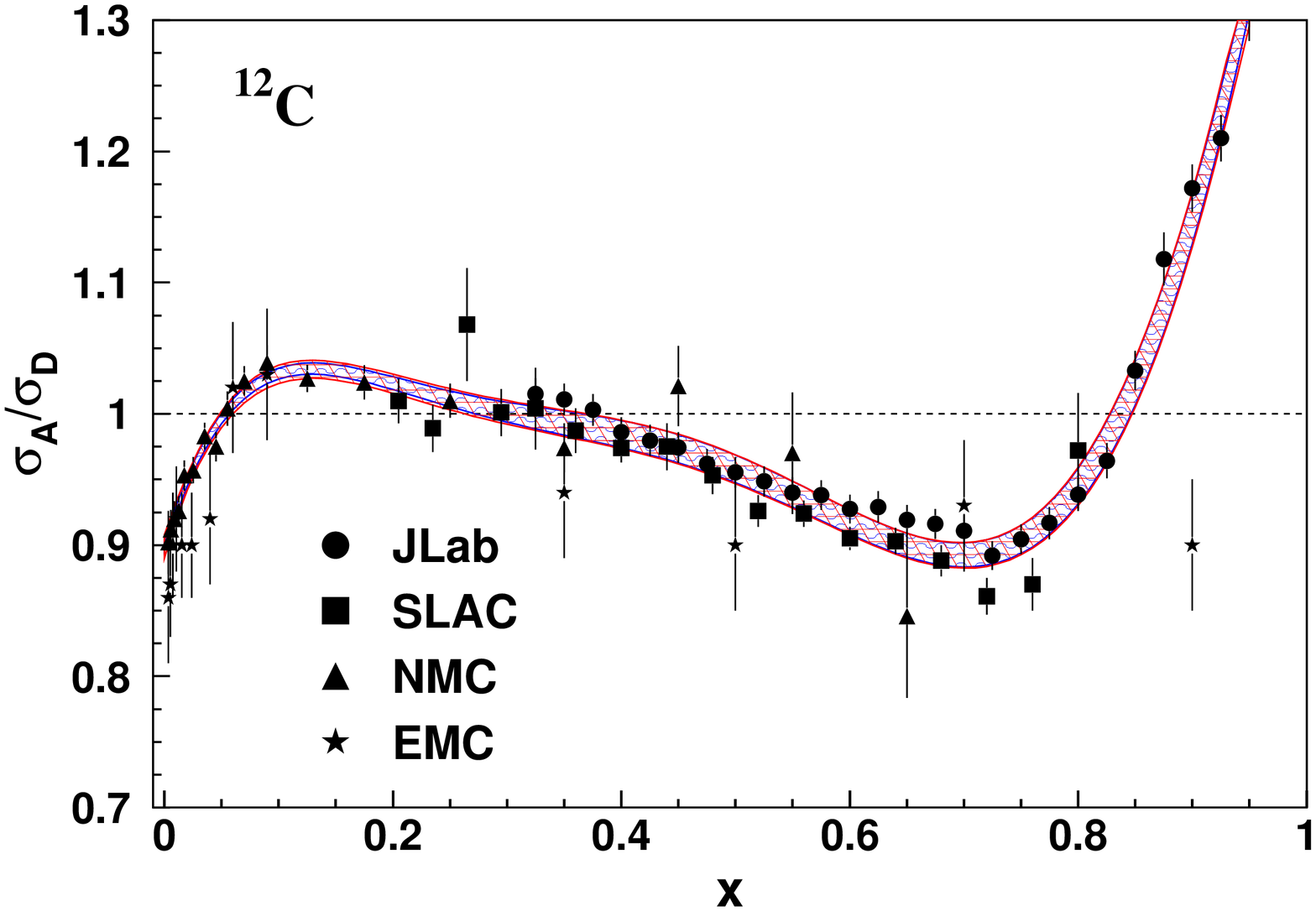}\\
\centering\includegraphics[width=\columnwidth]{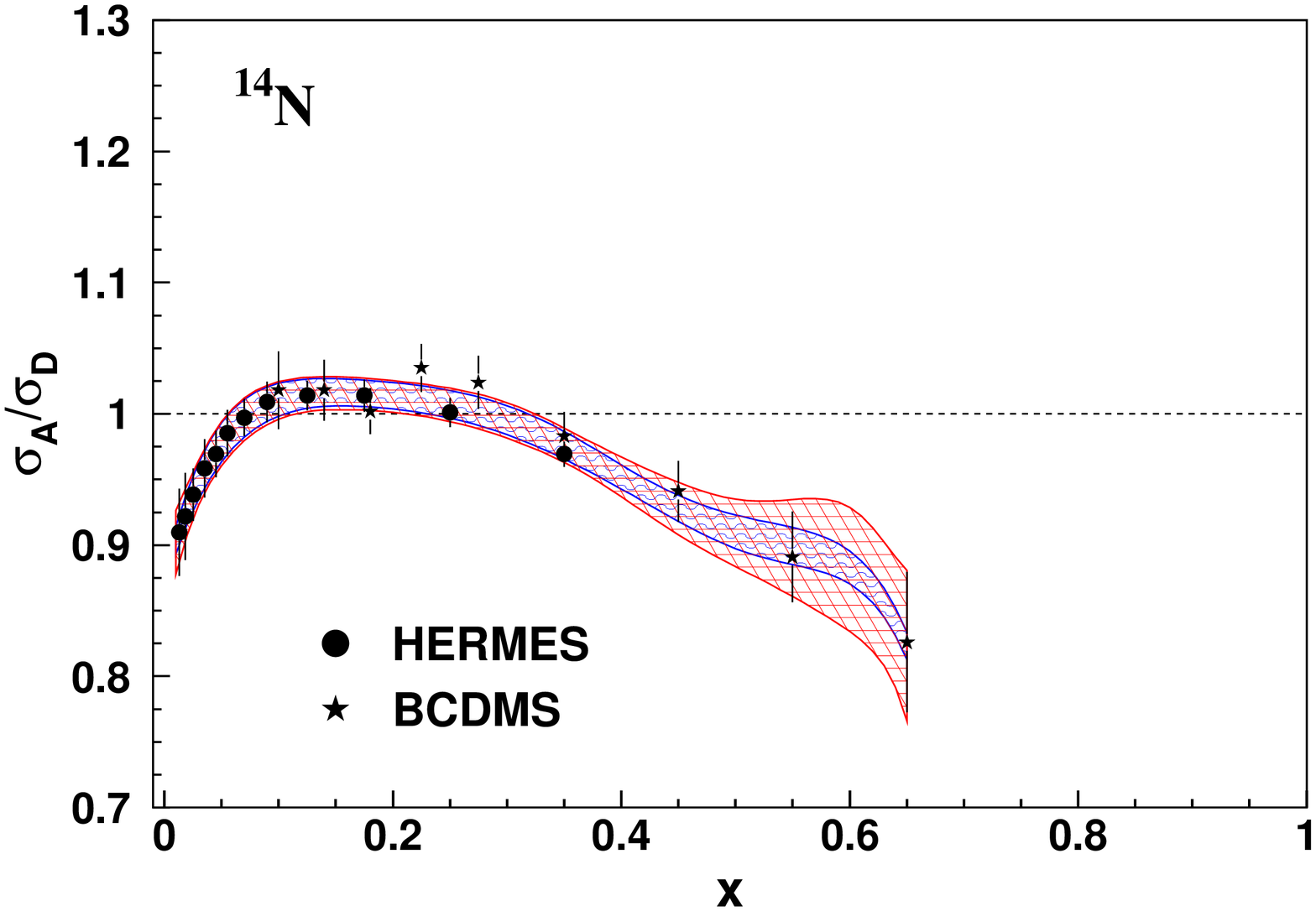} 
\caption{{\bf Top:} Ratios of the $^{12}$C to Deuterium DIS cross sections. 
The measurements shown are from JLab~\cite{Seely:2009gt} (circles), 
NMC~\cite{Arneodo:1995cs} (triangles), SLAC~\cite{Gomez:1993ri} (squares) and 
EMC~\cite{Arneodo:1989sy} (stars). 
{\bf Bottom:} Ratios of the 
$^{14}$N to Deuterium DIS cross sections. The measurements shown are from 
HERMES~\cite{hermes_he3_n:2003} (circles) and BCDMS~\cite{Bari:1985ga} 
(stars). In both panels the error bars 
represent the statistical and point-to-point systematic uncertainties added in 
quadrature. The bands show the fits to the combined data 
sets and their uncertainties when both the normalization and point-to-point 
errors are taken into account (see text for details).}
\label{fig:c_n}
\end{figure}

\begin{figure}[htbp]
\centering\includegraphics[width=\columnwidth]{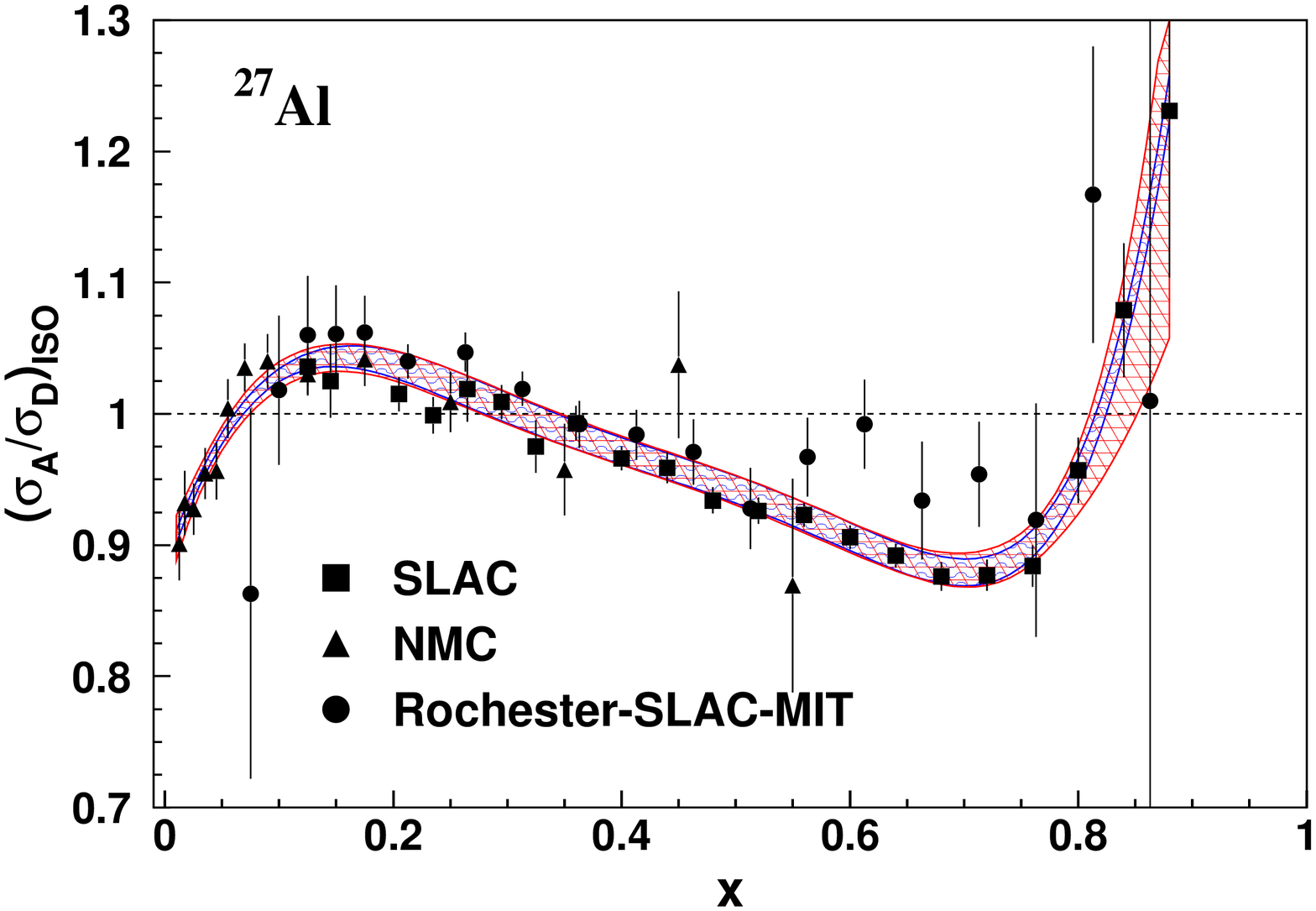}\\
\centering\includegraphics[width=\columnwidth]{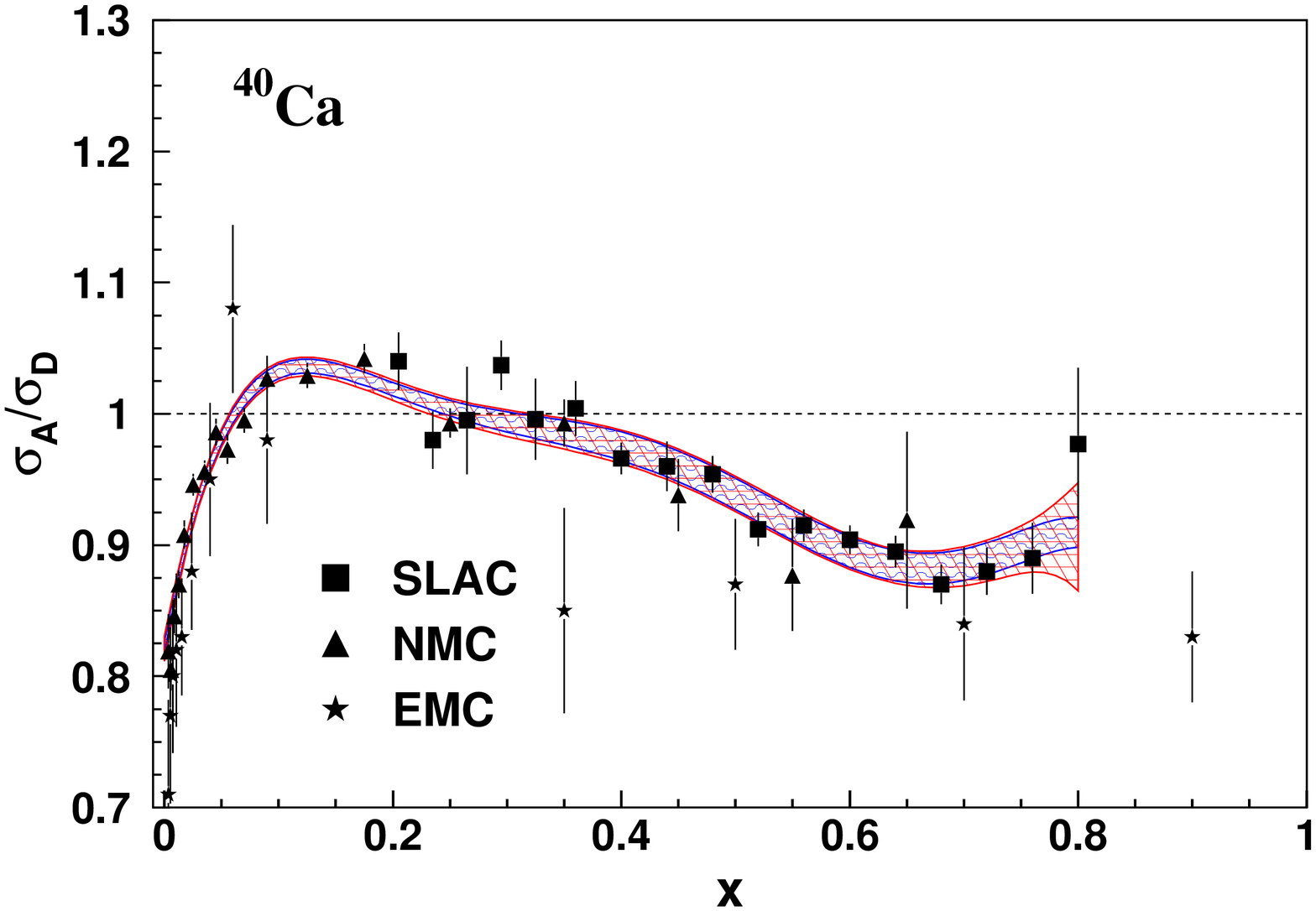} 
\caption{{\bf Top:} Ratios of the $^{27}$Al to Deuterium DIS cross sections. 
The measurements shown are from ROCHESTER-SLAC-MIT~\cite{Bodek:1983ec} (circles), NMC~\cite{Arneodo:1996rv,Amaudruz:1995tq} (triangles) and 
SLAC~\cite{Gomez:1993ri} (squares). 
{\bf Bottom:} Ratios of the 
$^{40}$Ca to Deuterium DIS cross sections. The measurements shown are from 
EMC~\cite{Arneodo:1989sy} (stars), NMC~\cite{Amaudruz:1995tq} 
(triangles) and SLAC~\cite{Gomez:1993ri} (squares). In both panels the error bars 
represent the statistical and point-to-point systematic uncertainties added in 
quadrature. The bands show the fits to the combined data 
sets and their uncertainties when both the normalization and point-to-point 
errors are taken into account (see text for details).}
\label{fig:al_ca}
\end{figure}

\begin{figure}[htbp]
\centering\includegraphics[width=\columnwidth]{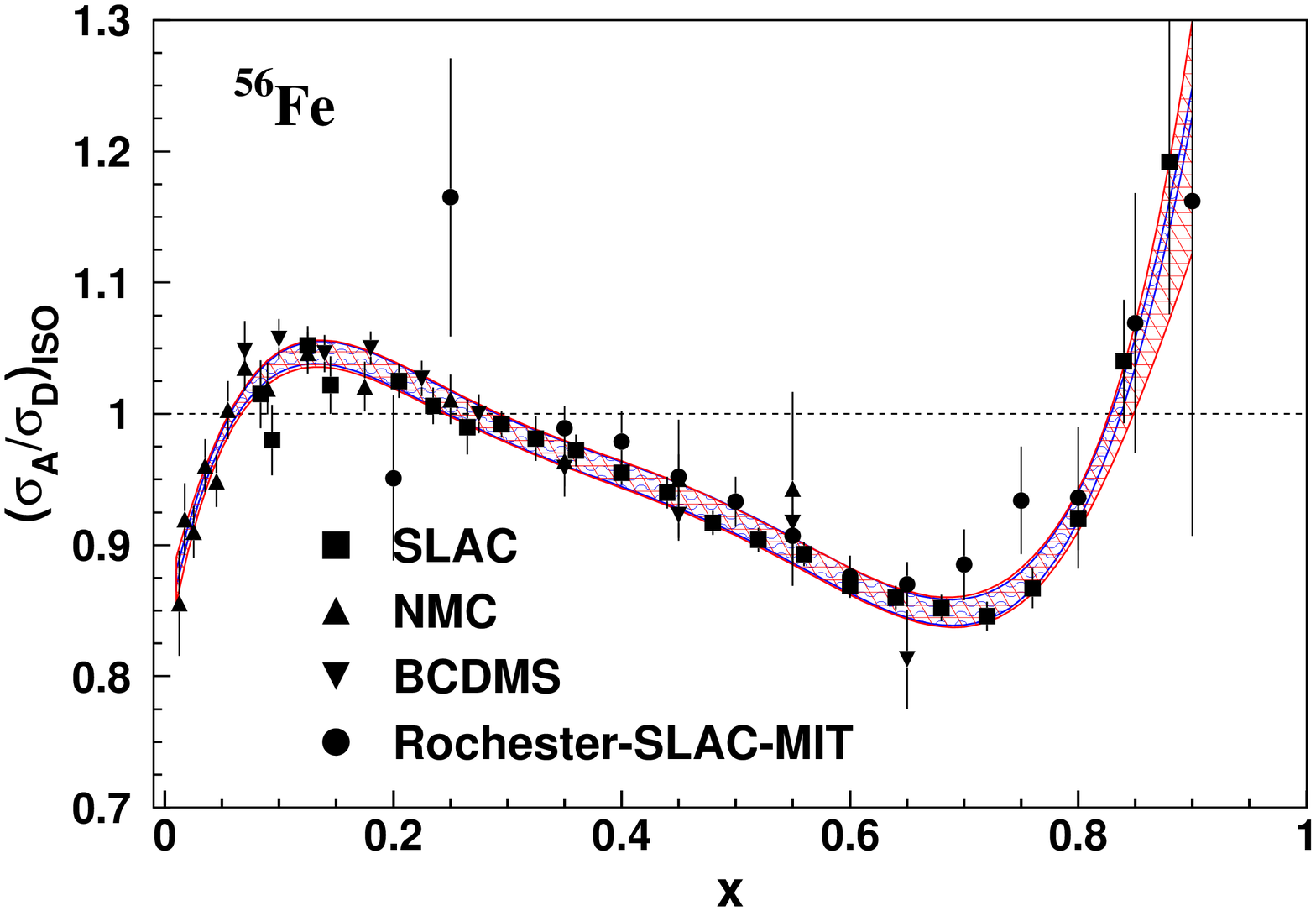}\\
\centering\includegraphics[width=\columnwidth]{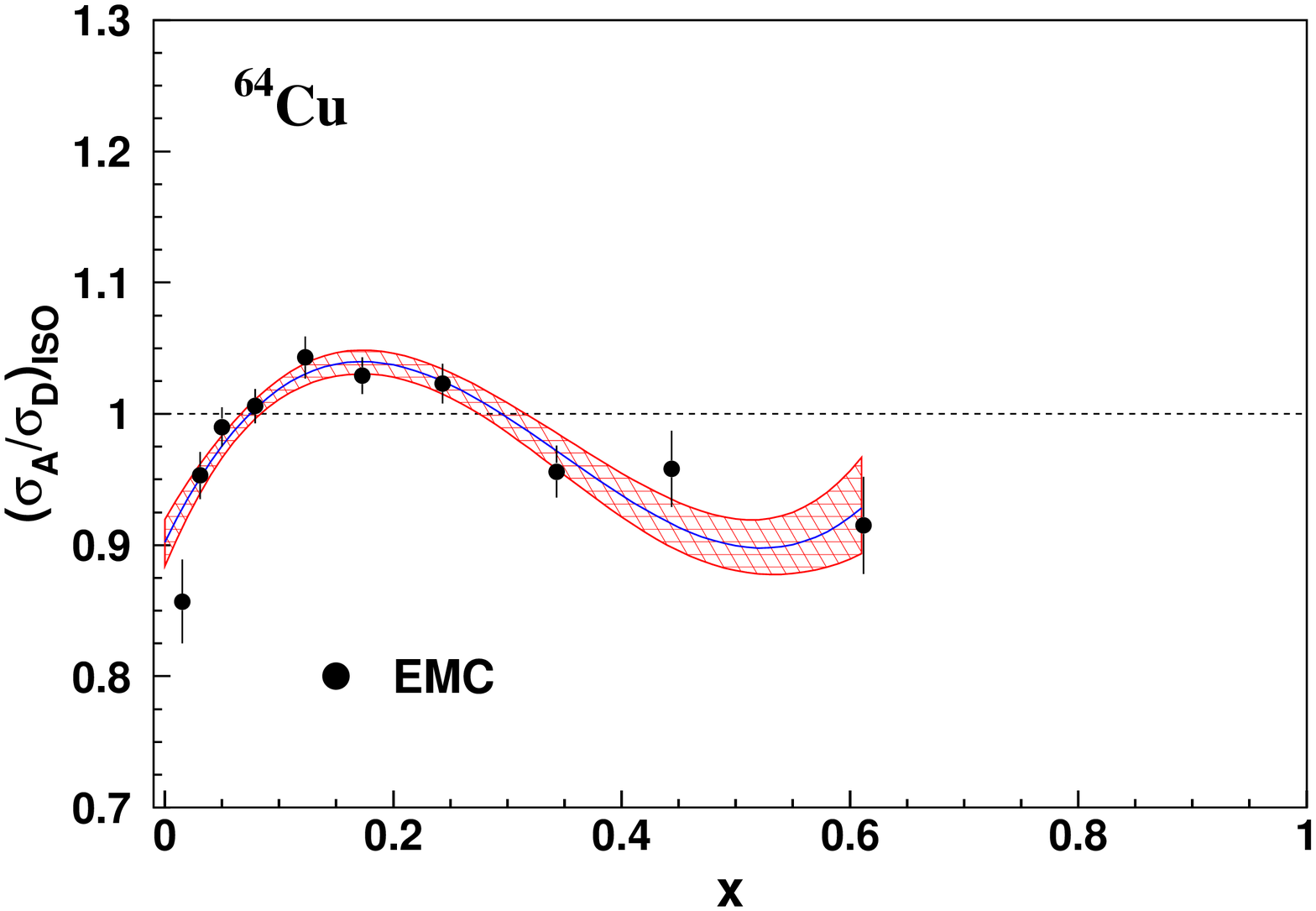} 
\caption{{\bf Top:} Ratios of the $^{56}$Fe to Deuterium DIS cross sections. 
The measurements shown are from ROCHESTER-SLAC-MIT~\cite{Bodek:1983qn} (circles), NMC~\cite{Arneodo:1996rv,Amaudruz:1995tq} (triangles), 
SLAC~\cite{Gomez:1993ri} (squares) and BCDMS~\cite{Benvenuti:1987az} (flipped triangles). 
{\bf Bottom:} Ratios of the 
$^{64}$Cu to Deuterium DIS cross sections. The measurements shown are from 
EMC~\cite{Ashman:1992kv} (circles). In both panels the error bars 
represent the statistical and point-to-point systematic uncertainties added in 
quadrature. The bands show the fits to the combined data 
sets and their uncertainties when both the normalization and point-to-point 
errors are taken into account (see text for details).}
\label{fig:fe_cu}
\end{figure}

\begin{figure}[htbp]
\centering\includegraphics[width=\columnwidth]{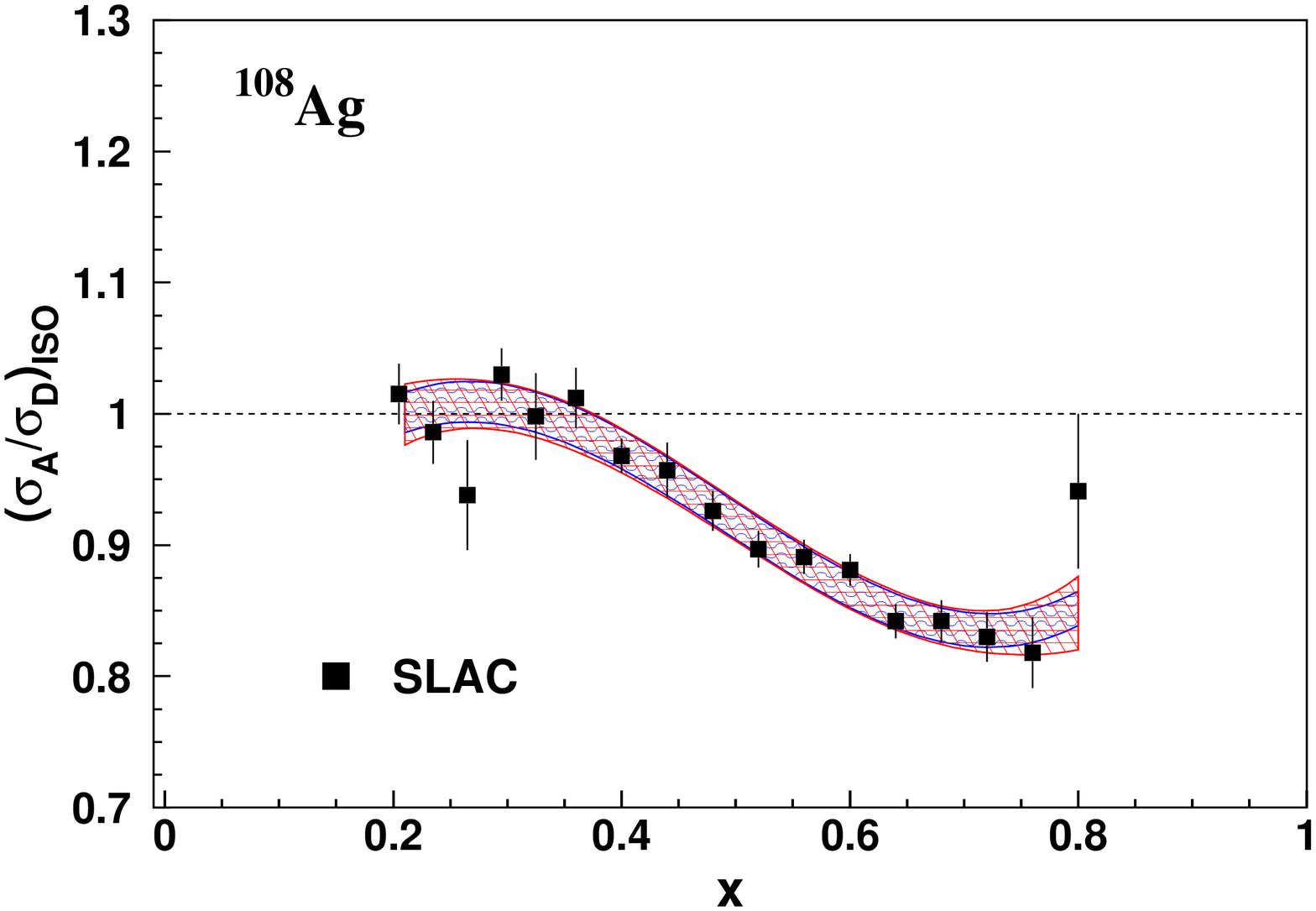}\\
\centering\includegraphics[width=\columnwidth]{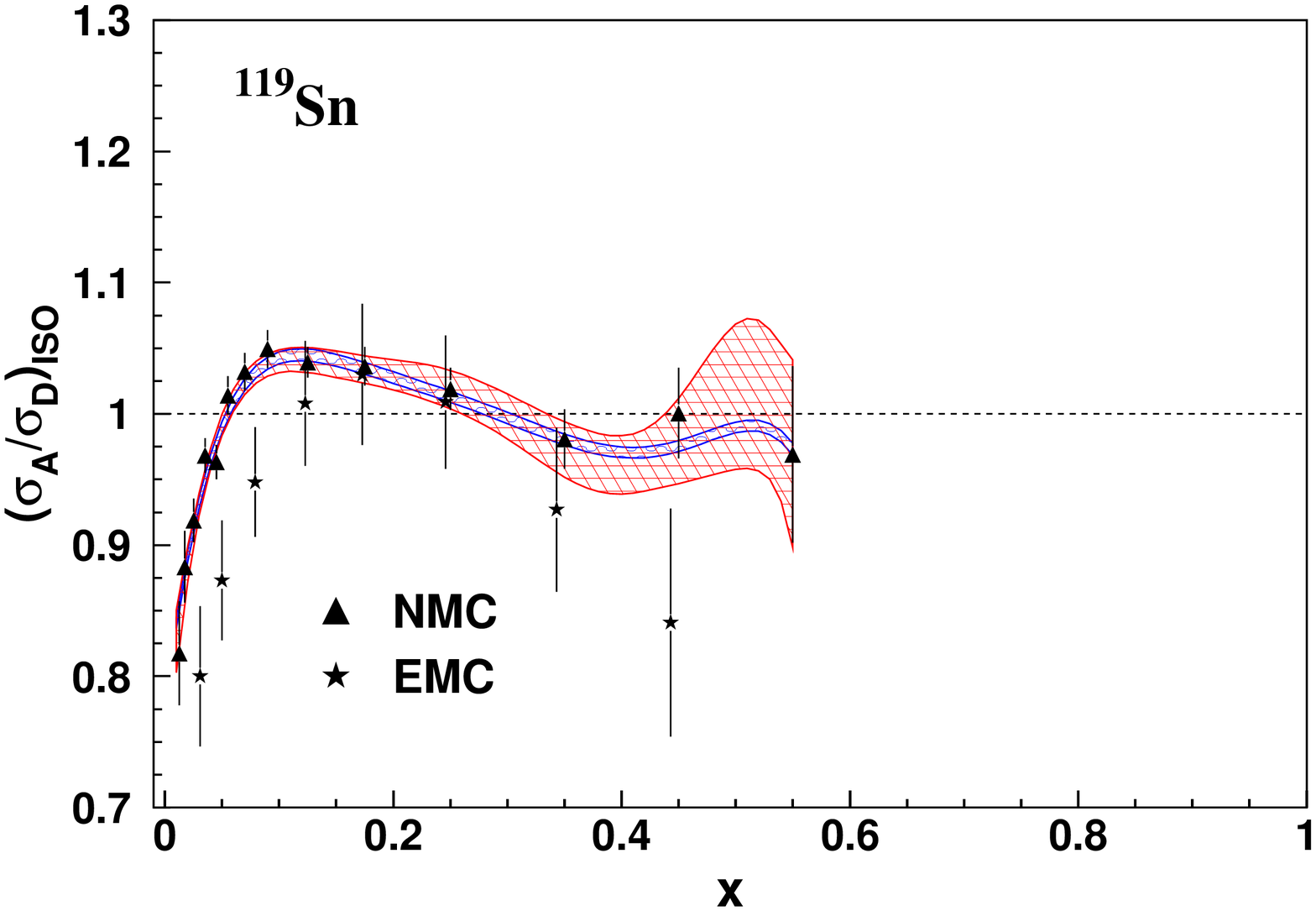} 
\caption{{\bf Top:} Ratios of the $^{108}$Ag to Deuterium DIS cross sections. 
The measurements shown are from SLAC~\cite{Gomez:1993ri} (squares). 
{\bf Bottom:} Ratios of the 
$^{119}$Sn to Deuterium DIS cross sections. The measurements shown are from 
NMC~\cite{Arneodo:1996rv,Amaudruz:1995tq} (triangles) and 
EMC~\cite{Ashman:1988bf} (stars). In both panels the error bars 
represent the statistical and point-to-point systematic uncertainties added in 
quadrature. The bands show the fits to the combined data 
sets and their uncertainties when both the normalization and point-to-point 
errors are taken into account (see text for details).}
\label{fig:ag_sn}
\end{figure}

\begin{figure}[tbp]
\centering\includegraphics[width=\columnwidth]{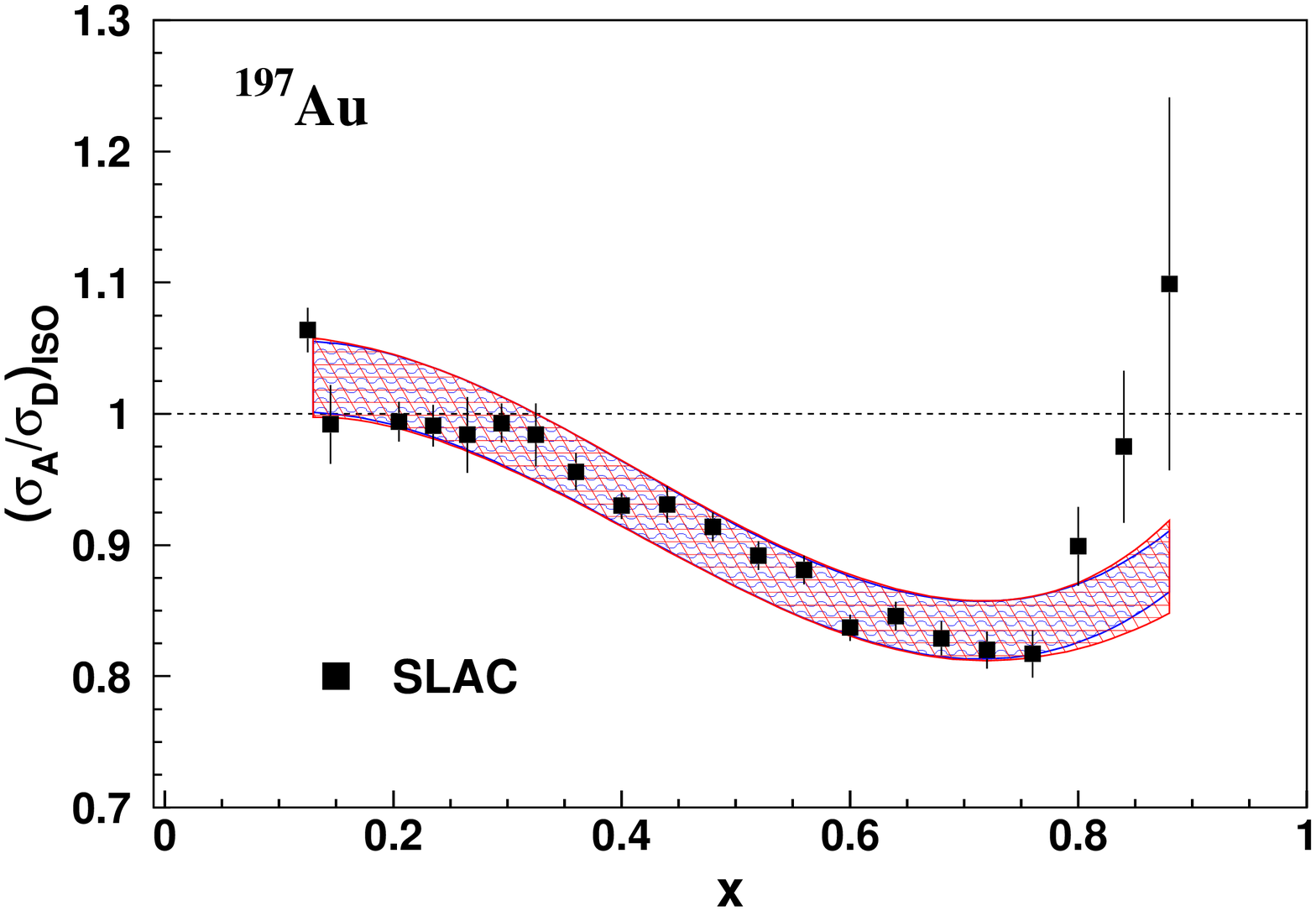}\\
\centering\includegraphics[width=\columnwidth]{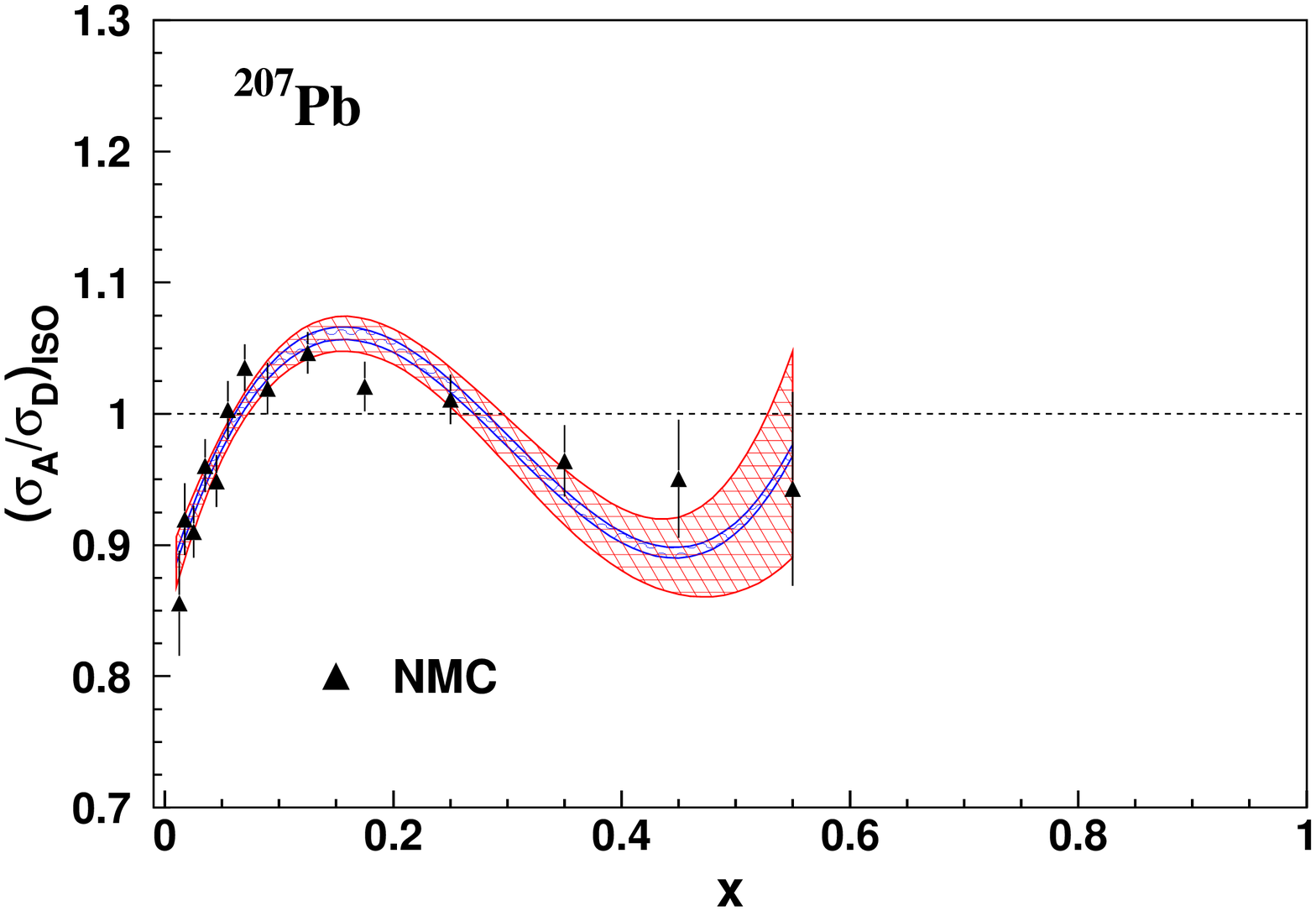} 
\caption{{\bf Top:} Ratios of the $^{197}$Au to Deuterium DIS cross sections. 
The measurements shown are from SLAC~\cite{Gomez:1993ri} (squares). 
{\bf Bottom:} Ratios of the 
$^{207}$Pb to Deuterium DIS cross sections. The measurements shown are from 
NMC~\cite{Arneodo:1996rv,Amaudruz:1995tq} (triangles). 
In both panels the error bars 
represent the statistical and point-to-point systematic uncertainties added in 
quadrature. The bands show the fits to the combined data 
sets and their uncertainties when both the normalization and point-to-point 
errors are taken into account (see text for details).}
\label{fig:au_pb}
\end{figure}

In Figs.~\ref{fig:he3_he4}--\ref{fig:au_pb} we show published measurements of
$(\sigma_{A}/A)/(\sigma_{D}/2)$ together with our fits for $^{3}$He, $^{4}$He,
$^{6}$Li, $^{9}$Be, $^{12}$C, $^{14}$N, $^{27}$Al, $^{40}$Ca, $^{56}$Fe,
$^{64}$Cu, $^{108}$Ag, $^{119}$Sn, $^{197}$Au and $^{207}$Pb. The data are
displayed as points with the point-to-point uncertainties as bars (statistical
and systematics added in quadrature) and our fits and their uncertainties are
shown as bands. The inner blue band shows the results of our analysis when only
the normalization uncertainty is included in the randomization procedure while
red hashed band includeds both the normalization and the
point-to-point uncertainties as explained earlier.  

The most precise experimental constraints on the nuclear structure functions
exist for $^{4}$He, $^{9}$Be, $^{12}$C, $^{27}$Al, $^{40}$Ca
and $^{56}$Fe. For these targets there are at least three independent data
sets in good agreement with each other, each covering a fairly wide range in
$x$.  For $^{9}$Be, $^{27}$Al and $^{56}$Fe the $(\sigma_{A}/A)/(\sigma_{D}/2)$
ratios for NMC have been obtained from their published results on $A/C$ and $C/D$
ratios. An upper limit on the overall uncertainty from our fits for the entire
$x$ range studied is approximately $2$\% for $^{4}$He, $^{9}$Be and $^{12}$C, 2.7\% for
$^{27}$Al for $x < 0.8$ growing to 12\% at $x=0.88$, 2$\%$ for $x < 0.75$
reaching 4.5\% at $x = 0.8$ for $^{40}$Ca and 1.7$\%$ for $x < 0.8$ increasing
to 7.3$\%$ at $x=0.9$ for $^{56}$Fe. $^{3}$He is also fairly well studied but
with very little overlap in $x$ between data sets. The $^{3}$He HERMES results as
published have been renormalized to agree with the NMC measurements on $^{4}$He
and nitrogen.  We decided to remove the normalization and use the HERMES ratios
as measured with a normalization uncertainty of 1.4\% as specified in their
publication. An upper limit on the overall uncertainty from our fits for
$^{3}$He is 2.5\%.

For $^{6}$Li, $^{14}$N, $^{64}$Cu, $^{119}$Sn and $^{207}$Pb there are no 
precision measurements beyond $x$ of $0.3-0.35$ and more data would be needed to study 
the EMC effect for these targets. For $^{108}$Ag and $^{197}$Au the SLAC 
measurements provide a fairly good mapping of the $x$ region extending from 
0.2 to 0.8 but there are no measurements in the shadowing and the 
antishadowing regions as well as in the large $x$ regime.

Of great interest has been the extraction of the ``size'' of the EMC effect 
from data. The procedure used in recent analyses involves the 
extraction of the slope $|dR_{EMC}/dx|$ from a linear fit to cross 
section ratios in the region of $0.35 < x <  0.7$~\cite{Arrington:2012ax}. This 
quantity is then studied in relation to $A$, the average/local nuclear 
density or the short range correlation factor in search for patterns that 
would shed light on the mechanisms by which the EMC effect may arise.
Typically the slopes are extracted from 
individual data sets and are then combined taking into account the individual 
uncertainties. We extracted $|dR_{EMC}/dx|$ for $^{3}$He, $^{4}$He, 
$^{9}$Be, $^{12}$C, $^{27}$Al, $^{40}$Ca, $^{56}$Fe, $^{108}$Ag and $^{197}$Au from 
our global fits and the results are shown in Table~\ref{tab:slopes}.
The slopes were determined by fitting the result of the global fit (evaluated at the $x$ values of the data used in the fit).
The uncertainty on the slope was determined using the experimental data's point-to-point uncertainties for each ratio.

\begin{table}[htb]
\tbl{Extraction of the EMC effect ``size''
, $|dR_{EMC}/dx|$, from our
global fits to data (see text for details). Results from
\cite{Arrington:2012ax} are also shown.}
{\begin{tabular}{c|c|c}
\hline 
EMC Effect Targets        & Slopes from this work           &  Slopes from Arrington {{\it et al.}}~\cite{Arrington:2012ax} \\ \hline
$^{3}He/D$ & 0.099 $\pm$ 0.027 & 0.070 $\pm$ 0.028 \\
$^{4}He/D$ & 0.222 $\pm$ 0.024 & 0.197 $\pm$ 0.025 \\
$^{9}Be/D$ & 0.266 $\pm$ 0.023 & 0.247 $\pm$ 0.023 \\
$^{12}C/D$ & 0.324 $\pm$ 0.022 & 0.292 $\pm$ 0.023 \\
$^{27}Al/D$ & 0.328  $\pm$ 0.031 & 0.325 $\pm$ 0.034 \\
$^{40}Ca/D$ & 0.360 $\pm$ 0.042 & 0.350 $\pm$ 0.047 \\
$^{56}Fe/D$ & 0.387 $\pm$ 0.026 & 0.388 $\pm$ 0.033 \\
$^{108}Ag/D$ & 0.496 $\pm$ 0.051 & 0.496 $\pm$ 0.052 \\
$^{197}Au/D$ & 0.393 $\pm$ 0.039 & 0.409 $\pm$ 0.040 \\
\hline 
\end{tabular} 
\label{tab:slopes} }
\end{table}

Our extracted values compare reasonably well with previously published results, 
the differences being no larger than 10$\%$ for all targets analyzed 
with the exception of $^{3}$He. For this target we record a slightly larger difference 
which originates from data displaying some sensitivity to the $x$ range chosen for fitting. 
The slope extraction from the global fit is 
less sensitive to the chosen $x$ range, with the result changing by no more than 2$\%$. 
This stability is automatically ensured by taking into account the 
experimental constraints at the boundaries of the EMC effect region. 
Generally an experimental fit of $(\sigma_{A}/A)/(\sigma_{D}/2)$ 
with $x$ can be used for extractions of quantities like $|dR_{EMC}/dx|$ if a very 
flexible fit form is employed to ``shape'' this dependence from combined 
data sets. In the end, such a fit should reflect the quantity/quality 
of available data and not the suitability of the fit function. Also the 
built-in assumption when using this method is that all uncertainties for 
all data sets have been accurately estimated by the respective collaborations. 
The advantage 
of obtaining such a fit from combined data sets resides in having a 
unified experimental description of the $(\sigma_{A}/A)/(\sigma_{D}/2)$ 
behavior with $x$.  

We conclude that three decades of experimental efforts concentrated on 
extracting $(\sigma_{A}/A)/(\sigma_{D}/2)$ have produced a large body of 
data, but that further measurements would still be useful for targets like $^{6}$Li, 
$^{14}$N, $^{64}$Cu, $^{119}$Sn and $^{207}$Pb, especially in the EMC effect 
region. Though data from SLAC mapped this $x$ region of interest for $^{108}$Ag 
and $^{197}$Au, additional experimental constraints would be useful. Presently 
there are no measurements for $^{108}$Ag and $^{197}$Au for $x < 0.3$. These 
measurements are needed to verify whether the $x$--dependence of nuclear medium 
modifications to nucleon structure is universal.
Additionally, the question of whether $R=\sigma_{L}/\sigma_{T}$ is modified in the 
nuclear medium still awaits a definitive answer.
\clearpage

\subsection{The Drell-Yan Reaction}\label{sec:drell_yan}

Deep inelastic scattering probes the charge weighted sum over all quark flavors, and in
that respect can be considered ``flavor agnostic''. On the other hand, the proton and pionic 
Drell--Yan reactions provide access to the nuclear modifications of anti--quark and valence 
flavor-dependent quark distributions. 
The importance of the Drell-Yan reaction, for its ability to
provide additional information on the mechanism responsible
for the EMC effect, was first noted in
Refs.~[\refcite{Ericson:1984vt,Bickerstaff:1985ax,Bickerstaff:1985da}].
The fundamental process of interest in this case is the annihilation of
a quark--antiquark pair, subsequent creation of a virtual photon and its decay into a lepton pair; 
$q + \bar{q} \rightarrow \gamma^* \rightarrow l^+ + l^-$

The cross section for the Drell-Yan process, $h+A \rightarrow \gamma^*(l^+l^-) +X$, in the quark parton picture
can be written,
\begin{equation}
\frac{d\sigma}{dx_1 dx_2} = \frac{4\pi \alpha^2}{9 M^2}\sum_i e^2_i\left[q^1_i(x_1)\bar{q}^2_i(x_2) +
\bar{q}^1_i(x_1)q^2_i(x_2) \right],
\label{eq:dy_firstorder}
\end{equation}
where $x_1$ and $x_2$ are the momentum fractions of the quarks in the beam and
target hadrons and $M$ is the invariant mass of the final two--lepton
system. While this leading order expression is too naive and requires
significant higher order corrections, calculations of these corrections are
tractable.

In the case in which the quarks in the hadron beam ($q^1$) are primarily at
large $x$, the first term in Eq.~\ref{eq:dy_firstorder} dominates and the
Drell-Yan reaction is sensitive to antiquark distributions ($\bar{q}^2$) in the
target. Hence, the Drell-Yan process can be used to provide information on the
modification of sea-quarks in the nuclear medium.

\begin{figure}[tbp]
\centering\includegraphics[width=\columnwidth,clip=true,trim=0 0 10mm 15mm]{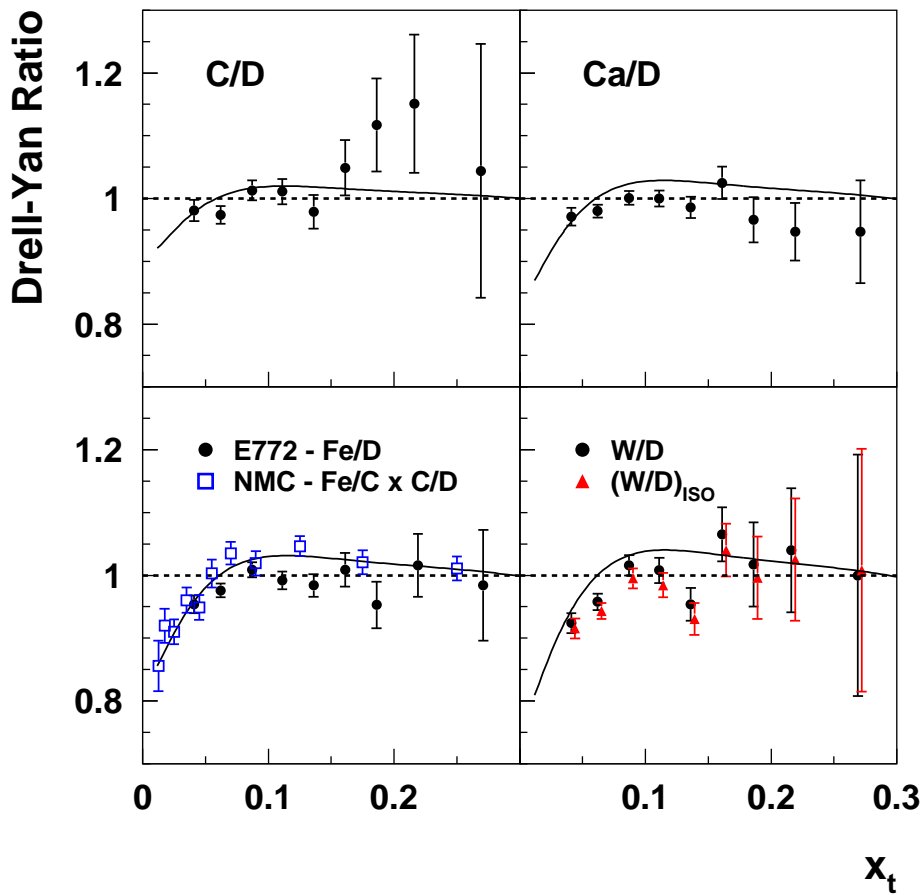}
\caption{Drell-Yan ratio of C, Ca, Fe, and W relative to deuterium
from Fermilab E772~\cite{Alde:1990im}.  Curves from the SLAC $A$-dependent fit
to the EMC effect in DIS~\cite{Gomez:1993ri} are also shown for 
comparison. Blue squares denote the EMC effect in iron extracted from 
NMC DIS ratios for Fe/C~\cite{Arneodo:1996rv} and C/D~\cite{Amaudruz:1995tq}. 
The bottom left panel (red triangles) also includes an isoscalar correction as 
discussed in the text.\label{fig:e772}}
\end{figure}

Results from Fermilab experiment 772~\cite{Alde:1990im}, in which the ratio of
Drell-Yan cross sections, (p-A)/(p-D), was measured for carbon, calcium,
iron and tungsten are shown in Fig.~\ref{fig:e772}. Over the measured range of
$x$ ($0.04<x<0.27$), there is little evidence for significant medium
modification of the anti-quark distributions in nuclei. This result is in
contrast with early explanations of the EMC effect in which much of the effect
was attributed to the presence of excess (virtual) pions in nuclei, and which
predicted large effects for the Drell-Yan ratio~\cite{Berger:1983jk}. Also of
note in this result is the apparent absence of any excess in the ratio in the
nominal antishadowing region, although this absence should be carefully
considered in the context of the $\sim$2\% normalization uncertainty associated with
the Drell-Yan ratios.

The published E772 results did not include any corrections to account for the fact that the Drell-Yan cross section
may differ for protons and neutrons, thus masking a potential nuclear dependence for $N \neq Z$ nuclei. As
described in Section~\ref{sec:formalism}, an isoscalar correction is typically applied to DIS measurements of the 
EMC effect to remove this ``trivial'' effect.  In the case of the nuclear dependence of the Drell-Yan cross section, 
there was an assumption that proton and neutron cross sections should be of similar magnitude. However, results from the subsequent
experiment 866 showed that this was definitively not the case~\cite{Hawker:1998ty}. In E866, the Drell-Yan
ratio $D/p$ was used to extract the light antiquark flavor asymmetry over an $x$-range similar to E772.
As an exercise, we have applied a naive isoscalar correction to the E772 tungsten data using a simple 
polynomial fit to the E866 data. The result of this correction is shown by the red triangles in 
Fig.~\ref{fig:e772} and is seen to decrease the ratio slightly, but by a very small amount, typically less than
1\%. However, it will be interesting to see if such an isoscalar correction is relevant for the next-generation
Drell-Yan experiment (E906~\cite{Isenhower:2001zz} at Fermilab) which will extend the available $x$ 
range to $\approx0.45$, where the nuclear effects may be significantly larger. 

In contrast to Drell-Yan measurements with proton beams, which provide access to the ``anti-quark EMC effect,''
pion Drell-Yan allows access to a potential valence quark flavor dependence, {\it{e.g.}}, a difference between 
up and down quark PDFs, $u_A(x)$ and $d_A(x)$. Assuming isospin symmetry in the pion PDFs ($u_{\pi^+}=d_{\pi^-}$,
$u_{\pi^-}=d_{\pi^+}$, $\bar{u}_{\pi^-}=\bar{d}_{\pi^+}$, and $\bar{u}_{\pi^+}=\bar{d}_{\pi^-}$), $A/D$ and charge
ratios for the pion-induced Drell-Yan reaction can be expressed
\begin{align}
\frac{\sigma^{DY}(\pi^+ + A)}{\sigma^{DY}(\pi^- + A)}& \approx \frac{d_A(x)}{4 u_A(x)},\\
\frac{\sigma^{DY}(\pi^- + A)}{\sigma^{DY}(\pi^- + D)}& \approx \frac{u_A(x)}{u_D(x)},\\
\frac{\sigma^{DY}(\pi^- + A)}{\sigma^{DY}(\pi^- + H)}& \approx \frac{u_A(x)}{u_p(x)},
\end{align}
where only the dominant terms in the cross section have been retained.

\begin{figure}[tbp]
\centering\includegraphics[width=\columnwidth,clip=true,trim=10mm 0 0 0]{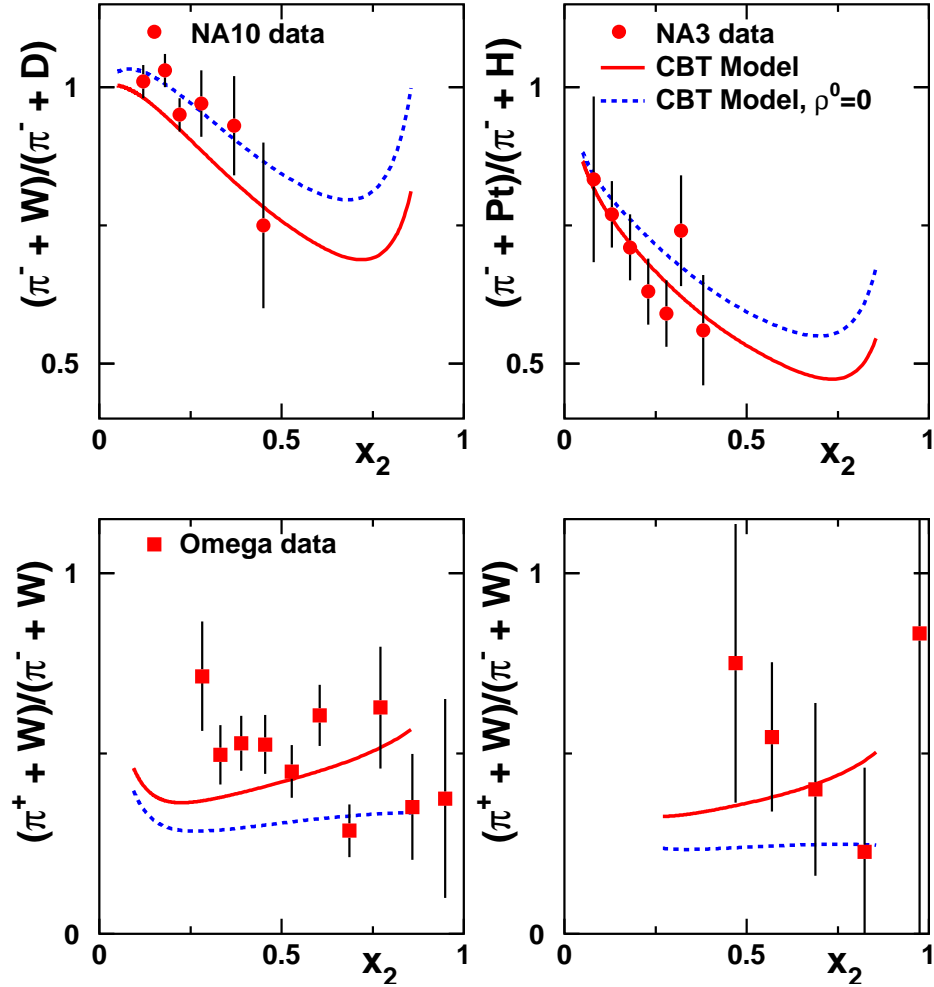}
\caption{Comparison of pionic Drell-Yan data from NA3~\cite{Badier:1981ci, Michelini:1981dt}, 
NA10~\cite{Bordalo:1987cs}, and the Omega collaborations~\cite{Corden:1980xf} to calculations from
Ref.~[\protect\refcite{Cloet:2009qs}]. The red curves denote calculations including flavor-dependent modifications of the
nuclear quark distribution functions, while blue curves have no flavor dependence. Figure taken from 
Ref.~[\citen{Dutta:2010pg}].
\label{fig:piondy}}
\end{figure}

At present, there is limited data available on the pionic Drell-Yan reaction (see Ref.~[\refcite{Chang:2013opa}] for 
an in-depth overview). In particular, the pionic Drell-Yan reaction from nuclei has been measured by the 
NA3~\cite{Badier:1981ci, Michelini:1981dt} and NA10~\cite{Bordalo:1987cs} experiments at CERN which used $\pi^-$ to 
measure the $Pt/H$ and $W/D$ ratios, respectively, while the Omega collaboration measured the $\pi^+/\pi^-$ ratio
from tungsten~\cite{Corden:1980xf}. These data are globally consistent with the overall nuclear dependence of 
quark distributions observed in DIS. More recently, the pionic Drell-Yan data has been examined in the context of
a model that predicts significant differences in the modification of up and down quark distributions 
in nuclei (see Fig.~\ref{fig:piondy}) and it was
observed that this model is slightly favored over one that includes no flavor dependence, but with limited significance
due to the relatively large uncertainties of the data~\cite{Dutta:2010pg}. Future measurements at 
COMPASS-II~\cite{Gautheron:2010wva} could provide increased precision in measurements of pionic Drell-Yan from nuclei 
and potentially provide unambiguous information regarding the flavor dependence of the EMC effect.

A possible flavor dependence of the EMC effect is particularly interesting in the context of the observed correlation
between the size of the EMC effect and the $a_2=\sigma_A/\sigma_D$ ratio measured for inclusive electron scattering at 
$x>1$. One explanation for this correlation is that the EMC effect is driven by high-momentum nucleons in the nucleus.
Since these high momentum nucleons should primarily come from correlated nucleon pairs, the $a_2$ ratios
serve as an indication of the relative probability to find these high-momentum nucleons. One implication of a connection
between the EMC effect and high momentum nucleons is that a flavor dependence of the EMC effect should be induced for
$N\neq Z$ nuclei~\cite{Sargsian:2012sm,Sargsian:2012gj}. For heavy nuclei with $N>Z$, a given proton is more likely to be 
found in a correlated (high-momentum) pair than a neutron. Since the proton contains two valence up quarks, one would 
expect up quarks to experience greater modification in those nuclei. It is worth noting that a similar flavor dependence 
was first predicted using a mean-field approach with no reference to short-range correlations~\cite{Cloet:2009qs,Cloet:2012td}, so 
observation of such a flavor dependence is no guarantee of the validity of the high-momentum nucleon explanation of the 
EMC effect. However, the failure to see such a flavor dependence would pose a challenge to this picture.

\section{Holistic View of Inclusive Electron Scattering}

In the quest to understand the EMC effect, experimentalists focused their efforts on deep inelastic
kinematics and would typically immediately cut their data so that
$Q^2 > 2$ and $W > 2$.     For the beam energies that were used, those
cuts immediately limited the data to $x < 1$ kinematics, yet
nature does not stop the cross sections at $x = 1$ as the $(e,e')$ cross sections as a function of $x$, as
defined herein, range from zero to $A$.   And while most of that range is highly dynamic, the $x > 1$
region has a plateau  whose magnitude has been shown to be rather $Q^2$ independent.

\begin{figure}[tbp]
\includegraphics[width=\linewidth]{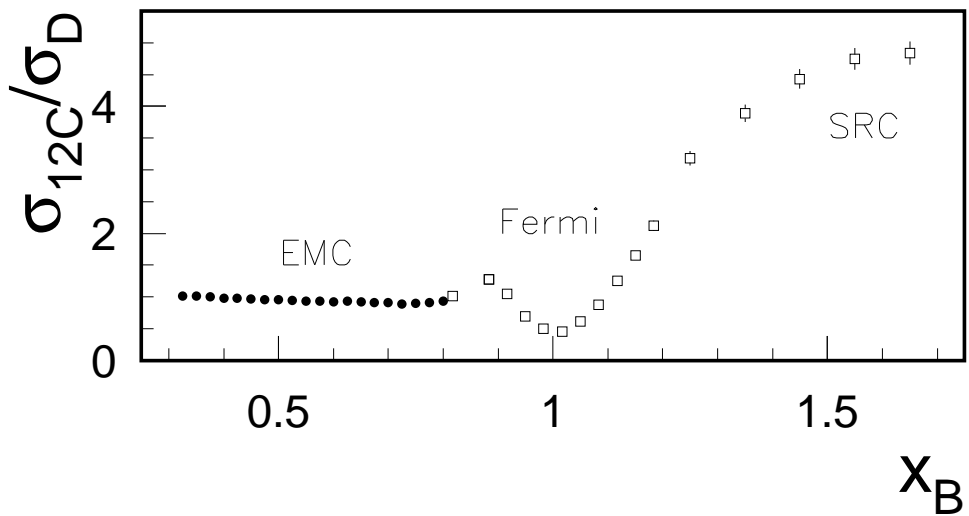}
\caption{Traditional plots of the EMC effect cut off at $x$ near one, but if one plots the full range of the
data, a high-$x$ plateau is observed beyond the Fermi momentum region.  The plot show deep inelastic scattering
data at a Q$^2$ of 4~[GeV/c]$^2$ from Seely~{\it{et al.}}~\cite{Seely:2009gt} and short-range 
data at a Q$^2$ of 2.7~[GeV/c]$^2$ from Fomin~{\it{el al.}}~\cite{Fomin:2011ng}.   
The magnitude of the short-range correlation plateau has been shown to be rather $Q^2$ independent 
just as the EMC effect slope is rather $Q^2$ independent.
This has led to speculation that the two effects are related either directly or indirectly~\cite{Weinstein:2010rt}.}
\label{fig:holistic}
\end{figure}

\begin{figure}[tbp]
\includegraphics[width=\linewidth]{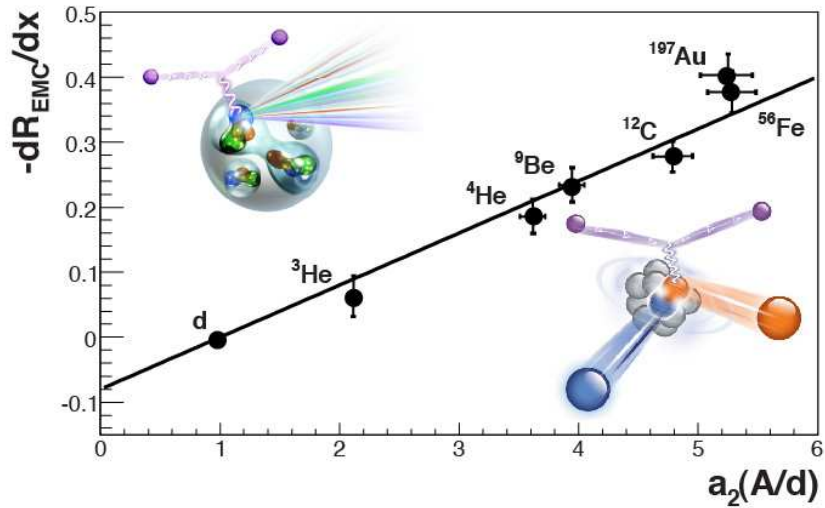}
\caption{Magnitude of the $x > 1$ plateau's plotted vs. the slope of the EMC effect is shown~\ref{Hen:2013oha}.   
Given that the EMC and SRC
analyses were independent, the correlation between the two results is striking.  The cartoons illustrate how different
these two reactions are with one being due to nucleon-nucleon knock-out and the other deep inelastic scattering.  
A logical question to ask from this experimental result is do the nucleon-nucleon correlations in the nucleus modify
parton distributions?  If so, it could explain the correlation between the effects.}
\label{fig:emc-src}
\end{figure}

In fact, once one plotted the entire $x$ range as in Fig.~\ref{fig:holistic}, phenomenologists could see that
the $x<1$ EMC dip and the $x>1$ SRC plateau~\cite{Higinbotham:2010ye} seemed to be
correlated when the data in those two regions were taken at similar $Q^2$.   
This holistic picture immediately lead to direct comparisons  of the slope of the EMC effect 
and magnitude of the SRC plateau.  The observed correlation,~\cite{Weinstein:2010rt,Hen:2012fm,Arrington:2012ax} 
as shown in Fig.~\ref{fig:emc-src}, from these completely independent experiments is quite striking.
  
The phenomenological observation
has reignited theoretical efforts into understanding the EMC effect~\cite{Higinbotham:2013hta}.   
In particular, there is new work in 
determining if a common underlying degree of
freedom is causing the correlation, such as high momentum initial states, binding effects, off-shell or
possibly the combination of all these nuclear effects, which when properly joined together, would explain
both regions (a deep inelastic theorist may take the opposite view and say that when effects of the nuclear medium
are properly taken into account this correlation can be understood as the modification of parton distributions).
Most beautifully, perhaps these two most extreme starting points actually meet for this one very special 
experimental result.

\section{Status of Theory}
The challenge of understanding the EMC effect~\cite{Aubert:1983xm} from a theoretical perspective has forced nuclear physicists to confront questions at the very heart of nuclear physics; namely, how do nuclei emerge from QCD and are there ascertainable remnants of this emergence in nuclear structure? After the more than 30 years since the discovery of the EMC effect the implications of QCD for nuclear structure are far from understood.

The immediate parton model interpretation of the EMC effect
is that the valence quarks inside a nucleus carry a smaller fraction of momentum than the valence quarks inside a free nucleon. 
The question remains however, what is the mechanism that causes this redistribution of quark momenta? Numerous explanations 
of the EMC effect have been proposed, e.g., 
nuclear binding~\cite{Akulinichev:1985ij,Akulinichev:1985xq,Akulinichev:1986gt,Akulinichev:1990su,Dunne:1985ks,Dunne:1985cn,Bickerstaff:1989ch,Benhar:1997vy}, pion excess in nuclei~\cite{Ericson:1983um,Berger:1983jk,Berger:1984na,Berger:1987er,Bickerstaff:1989ch}, multi-quark clusters~\cite{Jaffe:1982rr,Carlson:1983fs,Chemtob:1983zj,Clark:1985qu,Miller:1985un}, dynamical rescaling~\cite{Nachtmann:1983py,Close:1983tn}, medium modification~\cite{Bentz:2001vc,Mineo:2003vc,Smith:2003hu,Cloet:2005rt,Cloet:2006bq,Cloet:2009qs,Cloet:2012td} 
and short-range correlations~\cite{CiofiDegliAtti:1989eg,CiofidegliAtti:1990dh,Weinstein:2010rt,Arrington:2012ax,Frankfurt:2012qs,Hen:2012fm,Hen:2013oha}.
This section will canvas some of the more prominent ideas put forward to explain the EMC effect over the last several decades. 
Before doing so, however, we will briefly discuss a few salient aspects of DIS on nuclear targets which are pertinent to the EMC effect.

DIS on nuclear targets is potentially a rich area of experimental research that remains largely unexplored. For example, a nuclear target with total angular momentum $J$ has $2\,J+1$ independent structure functions in the Bjorken limit~\cite{Jaffe:1988up,Cloet:2006bq}, and thus far only the spin-averaged $F_2(x_A)$ structure function has been measured for nuclei with $A > 3$. Progress is being made however, with the HERMES data~\cite{Airapetian:2005cb} for the deuteron $b_1(x)$ structure function\footnote{An interpretation of the HERMES data, relevant to understanding the EMC effect, can be found in, e.g., Ref.~[\refcite{Miller:2013hla}].} and further measurements planned at Jefferson Lab~\cite{Chen:2011aa}. Measurement of the $2\,J+1$ structure functions for $A > 3$ nuclei, while challenging, could shed important light on the role of QCD in nuclear structure.

From a QCD perspective the various nuclei are bound states of an infinite number of quarks and gluons, defined by their valence quark content and discrete quantum numbers, in direct analogy to hadrons. The empirical fact that the valence quark content of nuclei can be replaced by their nucleon content in the classification of nuclei is already strong evidence for what may be called traditional nuclear physics, where the quarks and gluons of QCD are completely frozen inside the hadronic nuclear constituents and play no direct role in nuclear structure. In this picture QCD is of little relevance to nuclear structure beyond the nucleon--nucleon potential and the various hadronic matrix elements, e.g., nucleon form factors and parton distributions.

To confront the EMC effect theoretically one must determine the quark distributions of nuclei, which for quarks of flavor $q$ are defined by
\begin{align}
q_{A}\left(x_A\right) &= \int\frac{d\xi^-}{2\pi} e^{iP^+\,x_A\,\xi^-/A}
\left< A,P \left| \overline{\psi}_q(0)\,\gamma^+\,\psi_q(\xi^-)\right| A,P \right>,
\label{eq:nuclearpdf}
\end{align}
where $A$ labels the nucleus, $P$ its 4-momentum, $\psi_q$ is a quark field of flavor $q$ and $x_A$ is the Bjorken scaling variable.\footnote{The nuclear structure functions or quark distributions are often defined per-nucleon, in that case Eq.~\eqref{eq:nuclearpdf} and its corollaries would have an extra factor of $A^{-1}$.} 
In theoretical calculations it is customary to define the Bjorken scaling variable for a nuclear target as
\begin{align}
x_A \equiv A\,\frac{Q^2}{2\,P\cdot q} = A\,\frac{Q^2}{2\,M_A\,\nu} =  \frac{Q^2}{2\,\bar{M}_N\,\nu}
=  x_p\, \frac{M_p}{\bar{M}_N},
\label{eq:bjorkenxA}
\end{align}
where $A$ is the nucleon number, $q$ is the momentum transfer, $M_A$ is the mass of the nucleus,  $\bar{M}_N \equiv M_A/A$, $M_p$ is the proton mass and $x_p$ is the familiar scaling variable for the proton. The quark distributions (and structure functions) therefore have support on the domain $0 < x_A < A$. Note, in DIS experiments on nuclear targets the experimental structure functions are usually extracted from data as a function of the Bjorken scaling variable for the proton, $x_p$, therefore when comparing theory with experiment Eq.~\eqref{eq:bjorkenxA} should be used to replace $x_A$ with $x_p$~\cite{Hen:2013oha}.

For any explanation of the EMC effect to be credible the baryon number and momentum sum rules must remain satisfied, which for nuclear targets take the form:
\begin{align}
\label{eq:sumrulesbaryon}
&\int_0^A dx_A\, u_A^-(x_A) = 2\,Z + N, \hspace{5mm}
\int_0^A dx_A\, d_A^-(x_A) = Z + 2\,N,  \\
\label{eq:sumrulesmomentum}
&\int_0^A dx_A\, x_A\left[u_A^+(x_A) + d_A^+(x_A) + \ldots + g_A(x_A)\right] = Z + N = A,
\end{align}
where $Z$ and $N$ are the proton and neutron number, respectively; $g_A(x_A)$ is the nuclear gluon distribution function; and the plus and minus type quark distributions are defined by $q^{\pm}(x) = q(x) \pm \bar{q}(x)$.

All proposed explanations for the EMC effect discussed in the following sections provide a qualitative description of the data, at least in the region usually associated with the EMC effect, that is, $0.3 \lesssim x_A \lesssim 0.7$. Therefore an ability to distinguish between many of these various mechanisms will only be possible with new experiments that probe genuinely novel aspects of nuclear PDFs, important examples are their flavor dependence and spin structure. These new directions will be discussed from an experimental standpoint in Sect.~\ref{sec:future}, however in this section we will endeavor to highlight how this next generation of EMC type measurements will help to distinguish between the various mechanisms thought responsible for the EMC effect.

\subsection{Traditional Convolution Models}
\begin{figure}[tbp]
\centering\includegraphics[width=\columnwidth,clip=true,angle=0]{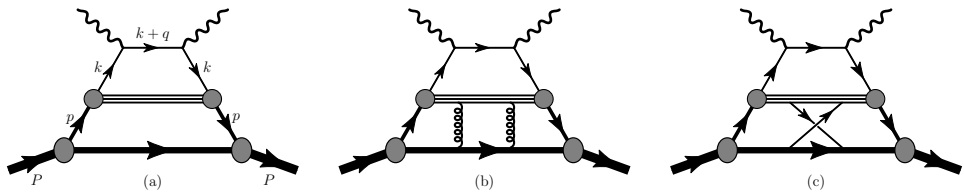}
\caption{Diagram (a) is a representation of the convolution formalism, where the virtual photon interacts with a quark confined inside a bound hadron. In the traditional convolution formalism diagrams of the type depicted in (b) and (c) are ignored, therefore the remnant of the struck hadron does not feel the nucleon medium during the interaction with the virtual photon. This is not the case in explanations of the EMC effect based on medium modification, which are discussed in Sect.~\ref{sec:medium_modification}. Diagrams like those illustrated in (b) and (c) do not vanish in the Bjorken limit~\cite{Jaffe:1985je}.}
\label{fig:convolution}
\end{figure}

From a traditional nuclear physics perspective it is natural to think of DIS on nuclear targets as a two step process; 
the virtual photon first scatters from a quark confined inside a hadronic nuclear constituent, 
giving the PDFs of the bound hadron; these PDFs are then combined with the lightcone distribution 
of the struck hadron inside the nucleus.
The total result gives the PDFs of the nuclear target. This process is illustrated in Fig.~\ref{fig:convolution}a and assumes an incoherent sum over the hadronic constituents; this formalism is often called the convolution model and ignores, e.g., the interaction of the structure hadron with the nuclear remnant as depicted in Figs.~\ref{fig:convolution}b and \ref{fig:convolution}c, which illustrate, respectively, gluon and quark exchange processes~\cite{Hoodbhoy:1986fn}.

The nuclear PDFs of Eq.~\eqref{eq:nuclearpdf} are represented in the convolution formalism by:
\begin{align}
q_{A}\left(x_A\right) &= \sum_{\alpha} \int_0^A dy_A \int_0^1 dz\ \delta\!\left(x_A - y_A\,z\right)\ f^{\alpha}_A(y_A)\ q_{\alpha}(z),
\label{eq:convolution}
\end{align}
where $\alpha$ is a sum over the hadronic nuclear constituents, e.g., nucleons, pions, deltas, etc. The quark distributions inside the bound hadrons $q_{\alpha}(z)$ are assumed to equal those of their free counterparts and $f^{\alpha}_A(y_A)$ is the lightcone momentum distribution (Fermi smearing function) of a hadron $\alpha$ in the nuclear target, which, in analogy to Eq.~\eqref{eq:nuclearpdf} has the formal definition
\begin{align}
f_{A}\left(y_A\right) &= \int\frac{d\xi^-}{2\pi} e^{iP^+\,y_A\,\xi^-/A}
\left< A,P \left| \overline{\psi}_\alpha(0)\,\gamma^+\,\psi_\alpha(\xi^-)\right| A,P \right>.
\label{eq:}
\end{align}
The scaling variable is given by $y_A = A\,\frac{p^+}{P^+}$ where $p^+$ is the plus-component of momentum for the bound hadron. For the baryon number and momentum sum rules of Eqs.~\eqref{eq:sumrulesbaryon} and \eqref{eq:sumrulesmomentum} to hold in the convolution framework, the Fermi smearing functions must satisfy
\begin{align}
\sum_{\alpha} \int_0^A dy_A\, f^{\alpha}_A(y_A) &= A, 
&
\sum_{\alpha} \int_0^A dy_A\, y_A \, f^{\alpha}_A(y_A) &= A.
\label{eq:fermisumrules}
\end{align}
What surprised many with the discovery of the EMC effect was that if the nucleus is assumed to consist only of nucleons ($\alpha = \text{protons}, \text{neutrons}$) then there exists no reasonable choice for $f^{p,n}_A(y_A)$ that can explain the EMC data~\cite{Miller:2001tg}. This observation changed our understanding of nuclear structure and its relation to QCD, because, prior to the discovery of the EMC effect, it was thought by many that quark and gluon degrees of freedom, characterized by the scale $\Lambda_{\text{QCD}} \simeq 250\,$MeV, would be unaffected by nuclear structure, which is typified by binding energies of $\sim\!10\,$MeV per-nucleon.

In nuclear structure pions are responsible for, \textit{inter alia}, the long range part of the nucleon--nucleon interaction. It is therefore natural to include pions in the sum of Eq.~\eqref{eq:convolution}. This idea implies an excess of pions in nuclei when compared to the free nucleon, and was one of the first explanations of the EMC effect~\cite{Ericson:1983um,Berger:1983jk,Berger:1984na,Berger:1987er,Bickerstaff:1989ch} grounded in traditional nuclear physics. The lightcone distribution of pions inside the nucleus $f^{\pi}_A(y_A)$ should peak at $y_A \sim m_\pi/\bar{M}_N$, whereas $f^{p,n}_A(y_A)$ has its peak at $y_A \sim 1$, therefore the introduction of pions into the convolution model shifts momentum from the valence quark region to the region where $x_A \lesssim 0.2$. In Ref.~[\refcite{Berger:1983jk}], for example, it was demonstrated that if pions carry 5\% of the nuclear lightcone momentum -- corresponding to an extra 0.13 pions per-nucleon -- then the qualitative features of the EMC effect in iron could be explained. 
An immediate consequence of the pion excess model is the prediction that there should be a significant enhancement of the sea-quark 
distributions in nuclei, however, as discussed in Sect.~IV the pion Drell-Yan experiments~\cite{Alde:1990im} at Fermilab 
in the late 1980s, and later charged pion electroproduction experiments at Jefferson Lab~\cite{Gaskell:2001fn,Gaskell:2001xr}, 
found no such enhancement.

The role of nuclear pions could potentially be clarified via measurements of the EMC effect in the spin-dependent structure function $g_{1A}$ (see Eq.~\eqref{eq:pemc} for a definition of the polarized EMC effect). Pions are spin zero and therefore the spin-dependent lightcone momentum distribution of pions inside a nucleus, $\Delta f^{\pi}_A(y_A)$, vanishes. A consequence therefore is that a naive pion excess model, based on Eq.~\eqref{eq:convolution}, should predict that the EMC effect in polarized structure functions would be much smaller than the original EMC effect, this is on contrast to the dynamical rescaling and medium modification approaches discussed in the following two sections.

\subsection{Dynamical rescaling}
Almost immediately after the discovery of the EMC effect it was observed that the per-nucleon $F_{2A}(x,Q^2)$ structure function of iron resembles the nucleon structure function $F_{2N}(x,Q^2)$, only at a larger value of $Q^2$, that is
\begin{align}
F_{2A}(x,Q^2) \simeq F_{2N}(x,\,\xi_A(Q^2)\,Q^2),
\label{eq:dynamical_rescaling}
\end{align}
where the rescaling factor satisfies $\xi_A(Q^2) > 1$. The properties of DGLAP evolution then produces the required quenching of the nuclear structure functions in the valence quark region but also a significant enhancement for $x \lesssim 0.1$, which, while seen in the original EMC data, disappeared in subsequent experiments.

This dynamical rescaling behavior was attributed to an increase in the confinement radius for quarks inside bound nucleons, as compared to their free counterparts; a consequence of the closely backed neighboring nucleons whose wave functions are often overlapping. A physical motivation for Eq.~\eqref{eq:dynamical_rescaling} can be obtained by noting that a factorization scale $\mu^2$ is associated with PDFs, which separates perturbative from nonperturbative physics. In perturbative QCD $\mu^2$ also provides an infrared cutoff for radiative gluons, therefore, an increased confinement radius for quarks inside bound nucleons implies $\mu_A^2 < \mu_N^2$. Assuming 
\begin{align}
q_A(x,Q^2=\mu_A^2) = q_N(x,Q^2=\mu_N^2),
\end{align}
leading-order DGLAP evolution then implies Eq.~\eqref{eq:dynamical_rescaling} where 
\begin{align}
\xi_A(Q^2) = \left[\frac{\mu_N^2}{\mu_A^2}\right]^{\alpha_s(\mu_N^2)/\alpha_s(Q^2)}.
\end{align}
If the confinement and factorization scales are related by
\begin{align}
\frac{r_A}{r_N} = \frac{\mu_N}{\mu_A},
\end{align}
then the value of $\xi \simeq 2$ at $Q^2 = 20\,$GeV$^2$ needed to qualitatively explain the EMC effect implies an increase in the confinement radius of about 15\%.

For polarized nuclear structure functions dynamical rescaling would predict a polarized EMC effect of a comparable size to the usual unpolarized EMC effect, therefore polarized DIS experiments on nuclear targets could help disentangle dynamical rescaling from alternative explanations of the EMC effect, like pion excess.

\subsection{Medium Modification \label{sec:medium_modification}}
The successes of more than 60 years of traditional nuclear physics teaches us that nuclei are composed largely of color single objects whose properties closely resemble those of free nucleons. These nucleons bound inside the nucleus are separated by distance scales similar to their actual size; as a consequence the wave functions of the bound nucleons are often overlapping. From a QCD perspective this could result in, e.g, quark exchange between nucleons~\cite{Hoodbhoy:1986fn} or the formation of hidden color configurations~\cite{Miller:2013hla}. It is therefore natural to conclude that the internal structure of all bound nucleons could be modified -- with respect to their free counterparts -- by the presence of the nuclear medium. This is known as medium modification. 

Aspects of medium modification are present in dynamical rescaling, associated with the change in confinement radius, however this idea was put on a firmer footing with the development of the quark-meson-coupling (QMC) model~\cite{Guichon:1987jp,Guichon:1995ue,Saito:2005rv}. In the QMC model nuclei are composed of bound nucleons
described by MIT bags, which are bound together by the exchange of mesons between the quarks of nearby nucleons. In its simplest form the model consists of an isoscalar--scalar ($\sigma$) and an isoscalar--vector ($\omega$) mean-field which self-consistently couples to the quarks inside the bound nucleons, causing a change in their internal quark structure. This model provides a self-consistent quark level description of the nuclear medium, giving the correct saturation properties of nuclear matter, and effective Skyrme forces in good agreement with other contemporary studies~\cite{Dutra:2012mb}.

\begin{figure}[tbp]
\subfloat{\centering\includegraphics[width=0.48\columnwidth,clip=true,angle=0]{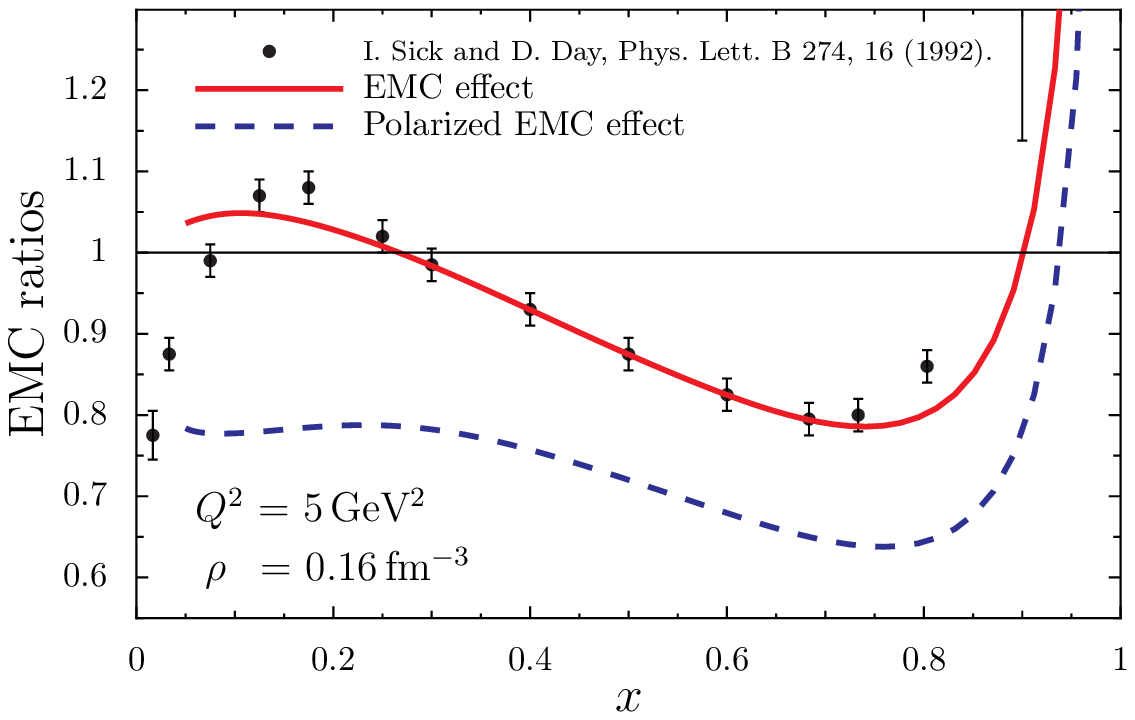}}
\subfloat{\centering\includegraphics[width=0.48\columnwidth,clip=true,angle=0]{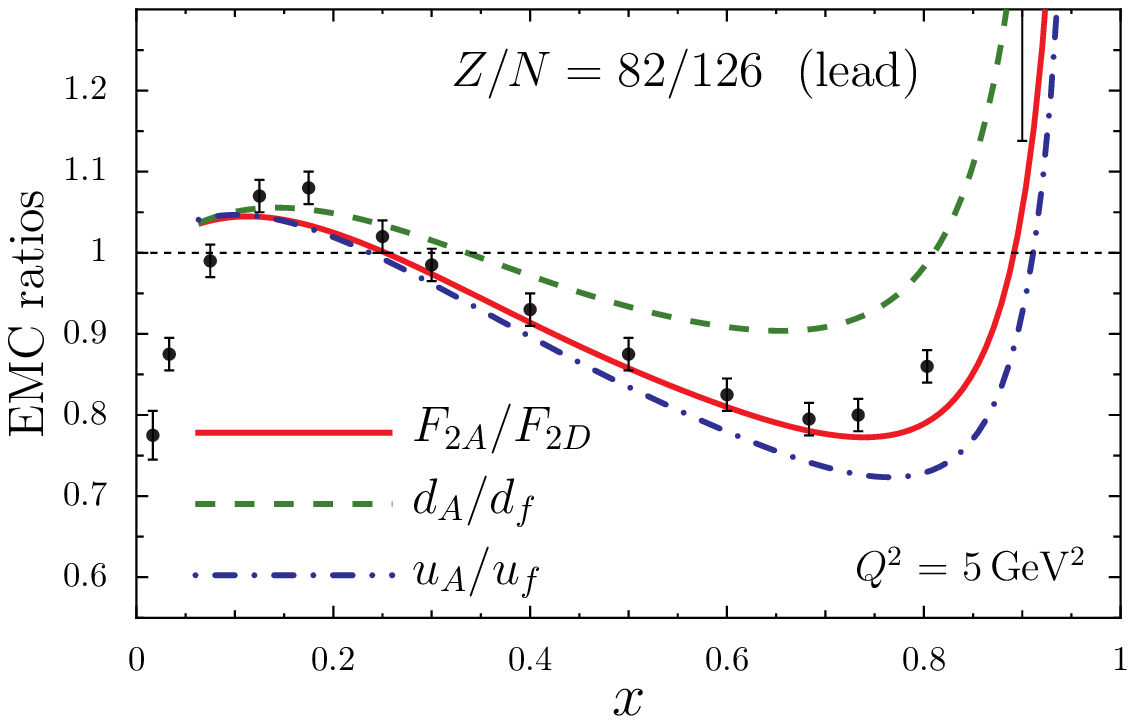}}
\caption{\textit{Left panel}: Results for the EMC (solid line) and polarized EMC (dashed line) effects in  nuclear matter from Refs.~[\protect\refcite{Cloet:2005rt,Cloet:2006bq}]. 
\textit{Right panel}: EMC effect (solid line) for nuclear matter with a $Z/N$ ratio equal to lead and this result split into quark flavors. The data in both figures is taken 
from Ref.~[\protect\refcite{Sick:1992pw}].}
\label{fig:emc_effects}
\end{figure}

The MIT bag model does not provide a covariant description of nucleon structure. This shortcoming has been addressed in Refs.~[\refcite{Bentz:2001vc,Mineo:2003vc,Cloet:2005rt,Cloet:2006bq,Cloet:2009qs,Cloet:2012td}] where a quark-level description for atomic nuclei has been constructed using the Nambu--Jona-Lasinio (NJL) model. This framework is similar in spirit to the QMC model in that the mean scalar and vector fields couple to the quarks inside the bound hadrons, thereby self-consistently modifying their quark structure. This model gives excellent results for the nucleon structure functions~\cite{Cloet:2005pp,Cloet:2007em} and provides a natural explanation for the EMC effect; the result for symmetric nuclear matter is illustrated as the solid line in the left panel of Fig.~\ref{fig:emc_effects}.
In this approach the EMC effect arises from an interplay between the mean scalar and vector fields in the nuclear medium. The scalar field reduces the dressed quark mass in-medium, $M^* < M$, which also reduces the nucleon mass, this tends to shift the nuclear PDFs to larger $x$ relative to the free nucleon. The mean vector field shifts the energy of all quarks in the medium, which produces a scale transform for a bound nucleon PDF given by
\begin{align}
q(x) &= \frac{p^+}{p^+ - V^+}\ \tilde{q}\left(\frac{p^+}{p^+ - V^+}\,x - \frac{V_q^+}{p^+ - V^+}\right),
\end{align}
where $\tilde{q}(x)$ is the nucleon PDF without the effect of the vector field, $p^+$ is the nucleon plus-component of momentum and $V^+$ is the plus-component of the mean vector field. The Fermi motion of the bound nucleons is included in a manner similar to Eq.~\eqref{eq:convolution}, except the sum is now over medium modified nucleons. The combination of these effects gives the results in Fig.~\ref{fig:emc_effects}.

This approach has also made predictions for the EMC effect in the spin-dependent nuclear structure function $g_{1A}(x_A)$. The polarized EMC effect can be defined by~\cite{Cloet:2005rt,Cloet:2006bq}
\begin{align}
\Delta R_{\text{EMC}} =\frac{g_{1A}}{g^{\text{naive}}_{1A}}
=\frac{g_{1A}}{P^{A}_{p}\,g_{1p} + P^{A}_{n}\,g_{1n}},
\label{eq:pemc}
\end{align}
where $g_{1p}$ and $g_{1n}$ are the free nucleon structure functions and $P^{A}_{p(n)}$ is the polarization of the protons (neutrons) in the nucleus $A$. The result for the polarized EMC effect -- for a polarized proton embedded in nuclear matter -- is given by the dashed line in the left panel of Fig.~\ref{fig:emc_effects}. The polarized EMC effect is found to be twice that of the usual EMC effect. A mean-field calculation using the chiral soliton model~\cite{Smith:2005ra} found a polarized EMC effect of comparable size to the spin-averaged EMC effect. The differences between the two models is attributed to the inclusion of antiquarks in the chiral soliton model. In these models the large polarized EMC effect is caused by a dramatic enhancement in the lower components of the quark wave functions in medium, which converts quark spin to quark orbital angular momentum.

This quark-level description of nuclei based on the NJL model has also made predictions for flavor dependence of the EMC effect for $N \neq Z$ nuclei~\cite{Cloet:2009qs,Cloet:2012td}. Predictions for this flavor dependence in nuclear matter with a $Z/N$ ratio equal to that of lead are illustrated in Figs.~\ref{fig:emc_effects}. In these calculations -- for $N>Z$ nuclei -- the isovector-vector mean field increases the binding of the $u$ quarks and decreases the binding for the $d$ quarks, which results in a significant flavor dependence for the EMC effect.

\subsection{Short-Range Correlations and Multi-quark Clusters}
The EMC effect originates because of the multiple nucleons that constitute nuclei; it is therefore natural to attempt to understand the EMC effect through features of the nucleon-nucleon potential. Sect.~3 introduced the recent observations of a series of papers~\cite{Weinstein:2010rt,Arrington:2012ax,Frankfurt:2012qs,Hen:2012fm,Hen:2013oha} which point out the strong linear correlation between the height of the $x_p > 1$  plateaus observed in inclusive quasi-elastic electron scattering and the slope of the EMC effect, for a large set of nuclear targets. The observed plateaus in the range $1.5 \lesssim x_p \lesssim  2.0$ are attributed to two-nucleon short-range correlations (SRCs) in the nuclear wave function. SRCs are characterized by high relative but low center-of-mass momentum, in comparison to the Fermi momentum, which is typically $k_F \simeq 200-270\,$MeV. Modern nuclear wave functions, e.g., determined using Green's function or variational Monte Carlo techniques, tend to have large high momentum components~\cite{Wiringa:2013ala}. For example, the momentum distributions of Ref.~[\refcite{Wiringa:2013ala}] give a probability of $\sim\!13\,$--$20$\% for a given nucleon to have a momentum greater than $270\,$MeV for nuclei with $4 \leqslant A \leqslant 12$. Detailed results are given in Tab.~\ref{tab:kf_probabilities}.

\begin{table}[b]
\addtolength{\tabcolsep}{5.0pt}
\centering
\tbl{Percentage probability that a proton or neutron will have a momentum greater than $270\,$MeV for various $A \leqslant 12$ nuclei; calculated from the momentum distributions of Ref.~[\protect\refcite{Wiringa:2013ala}].} 
{\begin{tabular}{l|cccccccc}
\hline\hline
             & ${}^2$H & ${}^3$H  & ${}^3$He & ${}^4$He & ${}^7$Li & ${}^9$Be & ${}^{11}$B & ${}^{12}$C  \\
\hline
proton (\%)  & 4.3     & 5.8      & 9.0      & 12.9    & 12.2     & 13.5     & 15.6     & 19.5    \\
neutron (\%) & 4.3     & 9.2      & 5.7      & 12.9    & 10.3     & 11.8     & 14.6     & 19.5    \\
\hline\hline
\end{tabular}}
\label{tab:kf_probabilities}
\end{table}

Explanations of the EMC effect based on SRCs also rely on the medium modification of the bound nucleon wave function~\cite{Sargsian:2012sm,Hen:2013oha}, where the degree of modification is proportional to the virtuality, $v = p^2 - M^2$, of the bound nucleon. In the previous section -- where the medium modification is induced by the mean fields -- each bound nucleon is modified at the level of a few percent.  If medium modification is caused by SRCs then an explanation of the EMC effect requires modifications many times larger (because far fewer nucleons undergo modification). With existing data it is not possible to distinguish between mean-field induced medium modification and that associated with SRCs~\cite{Hen:2013oha}. 

Knockout reactions have found that SRCs between $pn$ pairs are about a factor $20$ more likely than $pp$ or $nn$ SRCs. For $N > Z$ nuclei protons are therefore more likely to be associated with SRCs in the nuclear wave function. The SRC explanation of the EMC effect then implies that bound protons undergo larger medium modification than neutrons in nuclei with $N > Z$~\cite{Sargsian:2012sm}, such as lead. Therefore, since $u$ quarks dominate the proton, the up quark distribution in $N > Z$ nuclei should have a larger EMC effect than the down quark distribution. This flavor dependence of the EMC effect is similar to that induced by an isovector--vector mean field discussed in the previous section. As such, measurements of the flavor dependence of the EMC effect are unlikely to distinguish between the mean-field and SRC mechanisms. Measurements of the polarized EMC effect may however help unravel these two effects. For example, the proton spin contributes approximately $87$\%~\cite{Wiringa:2013ala} of the total angular momentum for the ${}^7$Li nucleus and of this $87$\% only $2.6$\% is from nucleons with momentum above $270\,$MeV; which is a factor $4$ less than the spin-averaged case (see Tab.~\ref{tab:kf_probabilities}). Mean-field induced medium modification produces an EMC effect and polarized EMC effect of comparable size for ${}^7$Li~\cite{Cloet:2006bq}, for SRCs to produce a similar result the medium modification of the bound $g_{1p}$ structure function must be $4$ times that for the spin averaged structure. Therefore, compared with mean-field induced medium modifications, comparable polarized and spin-averaged EMC effects require SRC induced medium modifications in spin structure functions approximately $20$ times larger than the mean-field approach. Therefore, comparable polarized and spin-averaged EMC effects appears less likely if SRCs are the mechanism responsible.

Two medium modification mechanisms that fit naturally with SRCs are quark exchange between nucleons~\cite{Hoodbhoy:1986fn} and the formation of multiquark ($6,\,9,\,12,\ldots$) clusters~\cite{Jaffe:1982rr,Carlson:1983fs,Chemtob:1983zj,Clark:1985qu,Miller:1985un}. Both of these ideas have been extensively discussed since the discovery of the EMC effect. In Ref.~[\refcite{Hoodbhoy:1986fn}] it was demonstrated that quark exchange between overlapping nucleons may explain a substantial portion of the EMC effect. Multiquark clusters shift valence quark momenta to $x_A > 1$, momentum conservation then demands a depletion in the EMC region. To explain the EMC effect in iron a six-quark bag probability of $20$--$30$\% is needed~\cite{Carlson:1983fs,Miller:1985un}, which is consistent with the probability of high momentum protons or neutrons in iron~\cite{Sargsian:2012sm}.

The slope of the EMC effect is not only correlated with the SRC plateaus, in  Refs.~[\refcite{Benhar:2012nj,Arrington:2012ax}] it was pointed out that there also exists a strong linear correlation with the mean removal energy for nucleons in nuclei. The mean removal energy is sensitive to the global properties of the nuclear wave function, not just the short distance piece associated with SRCs. In Refs.~[\refcite{Benhar:2012nj,Arrington:2012ax}] the mean removal energy was determined using the Koltun sum rule~\cite{Koltun:1974zz} with various approximations for the nuclear spectral functions. Although a correlation exists, direct calculations of the EMC effect incorporating these type of binding effects are generally thought to not fully explain the EMC effect~\cite{Hen:2013oha}.

\section{Future Directions \label{sec:future}}

Interest in the EMC effect remains high despite
the passing of the thirtieth anniversary of its original discovery.  The recent observation of the 
correlation of the EMC effect with the Short Range Correlation ``plateau'' has created a flurry of 
activity and led to plans for several new experiments. The EMC-SRC connection, however, is not the only
outstanding issue to be addressed with regard to the EMC effect.  Below we list some of the 
key topics that should be investigated in order to obtain a complete
picture of the origins of the EMC effect.  Experiments are already planned in many of these areas,
however, there are some holes that remain to be filled.

\subsection{Elucidation of the EMC-SRC Connection}

Exploration of the apparent connection between the EMC effect and Short Range Correlations is clearly
one of the highest priorities for future measurements. One of the most 
straightforward ways to probe this connection is to add to the database of EMC ``slopes'' and 
SRC ``plateaus'' already compiled~\cite{Weinstein:2010rt, Hen:2012fm, Arrington:2012ax}. In particular, 
the addition of nuclei with unique clustering structure (similar to beryllium), precision data
on the $^{40}$Ca and $^{48}$Ca isotopes, and targets with the largest practical $N/Z$ ratio
are particularly attractive. Such measurements are planned after completion of the Jefferson Lab
12~GeV Upgrade, and at least some of these data should be one of the earlier results of that 
program~\cite{E12_06_105,E12_10_008}.
It is worth noting that it would be helpful if these experiments could provide improved precision
for existing heavy target data, as well as better coverage of the anti-shadowing region for these
heavy targets.


A more direct study of the EMC-SRC connection would be to explore measurements of the 
inclusive $F_2$ structure function for a very high momentum nucleon in the nucleus.  An initial
attempt at such a measurement was reported by the CLAS collaboration at 
Jefferson Lab~\cite{Klimenko:2005zz}. In this case the D$(e,e'p_s)$X reaction was used to measure inelastic 
scattering from a bound neutron in the deuteron, where the initial nucleon momentum was determined by tagging the 
spectator proton. The inelastic structure function, $F_2(x)$, was measured as a function of the struck
neutron virtuality, however the results were somewhat ambiguous due to the relatively low beam energy available
(6 GeV) and apparent contributions from final state interactions.  

An updated version of this measurement will be performed as part of the Jefferson Lab 12 GeV 
program~\cite{E12_11_107}. In this case, the higher beam energy should aid in the interpretability
of the measurement. In addition, a broader kinematic coverage will help in the understanding of
final state interaction effects.

The next-generation version of this experiment would of course involve using a heavier
nucleus, such as $^4$He, for which the EMC effect is larger. It is also worth mentioning  that studies
of the EMC effect as a function of nucleon virtuality are ongoing, using existing data from Jefferson 
Lab Hall B, as part of the so-called ``Data-Mining'' initiative, in which large data sets from
separate experiments have been recast in a common framework to facilitate studies of 
nuclear-dependent observables~\cite{ODU_data_mining}.

\subsection{Flavor Dependence of the EMC Effect}

While the notion that the EMC effect could depend on valence quark flavor has always been entertained,
only recently have concrete predictions of the possible flavor dependence become available. One prediction of 
this flavor dependence~\cite{Cloet:2009qs,Cloet:2012td} results from the interaction of quarks with an
isovector--vector mean field and predicts a significant difference between the nuclear modification of
$u$ and $d$ quark distributions in $N \neq Z$ nuclei. More recently, it has been suggested that $u$ 
quarks will experience a larger nuclear modification in $N>Z$ nuclei if the EMC effect is driven by very high 
momentum nucleons 
resulting from Short Range Correlations~\cite{Sargsian:2012sm,Sargsian:2012gj}. The observation (or not) of such a flavor
dependence would place rigorous constraints on models that purport to describe the EMC effect and 
would provide perhaps one of the more exciting pieces of new information with regards to nuclear
PDFs.

Fortunately, there are several experimental avenues available for accessing the flavor dependence
of the EMC effect. Pionic Drell-Yan has already been discussed in Section~\ref{sec:drell_yan} and 
the COMPASS-II experiment~\cite{Gautheron:2010wva} at CERN should be able to provide information on 
this quantity in the near future.
Electron scattering also provides access to the flavor dependence via Parity Violating Deep
Inelastic Scattering (PVDIS) on heavy targets with $N>Z$ (Au or Pb for example) and such measurements
are planned as part of the Jefferson Lab 12~GeV program~\cite{SOLID_pvdis}, although they are not part of the 
first-generation of experiments planned for after completion of the upgrade, so the timeline for execution
is uncertain. The predicted size of the effect in PVDIS could be as large as 5\%  and should be within 
the experimental capabilities of state-of-the-art PV experiments. Additionally, Semi-inclusive Deep 
Inelastic Scattering (SIDIS) has been proposed as another mechanism by which one could ``flavor-tag'' the 
EMC effect. In this case the expected size of the relevant observable could be as large as 10\%, although 
quark hadronization effects in the nuclear medium could potentially cloud the interpretation of such an experiment. 
Finally, $W$ boson production in $p$--$A$ collisions at RHIC or LHC is another suggested avenue of 
investigation~\cite{Chang:2011ra}. Given the plethora of experimental observables available, it seems certain
that information regarding the flavor dependence of the EMC effect will become accessible within the next several
years.

\subsection{Nuclear Modification of Sea Quark Distributions}

At present, information concerning the modification of sea-quark distributions is limited to that provided
by the $p$-$A$ Drell-Yan reaction, i.e., the results from E772~\cite{Alde:1990im}. The present experimental results 
are limited to $x<0.3$ and the Drell-Yan $A/D$ ratios are more or less consistent with ratios from 
inclusive electron scattering, although there is an apparent lack of an enhancement of the ratio in the anti-shadowing 
region (although the significance of this is debatable given the relative normalization uncertainties of the Drell-Yan and DIS
data).

Data from the ongoing E906 experiment~\cite{Isenhower:2001zz} at Fermilab should prove very interesting, as 
this experiment will extend the $x$ coverage of the earlier E772 results up to $x \approx 0.45$, where the ratio of cross
sections becomes appreciably suppressed in DIS.

It has also been suggested that Semi-inclusive DIS could be used to probe the sea-quark distributions
in nuclei~\cite{Lu:2006xr}. While charged pion production in SIDIS has been shown to be sensitive to the 
valence quark flavor dependence, ratios of $\bar{p}/p$ and $\bar{\Lambda}/\Lambda$ display very different
behaviors for models with different assumptions concerning sea-quark distributions in nuclei. Unfortunately,
the Berger criterion for factorization in SIDIS~\cite{Berger:1987zu,Mulders:2000jt} suggests that such ratios 
would not likely be cleanly interpretable at the maximum beam energies accessible at Jefferson Lab, and would likely 
need to be measured at either a higher energy fixed target accelerator, or at a facility like the Electron-Ion Collider 
(EIC) currently under conceptual development at both Jefferson Lab and RHIC. As with semi-inclusive pion production, 
concerns with regards to effects from quark hadronization would also apply in this case.

\subsection{Polarized EMC Effect}

The experimental and theoretical emphasis has until recently focused on modifications of the unpolarized
quark distributions ($u(x), d(x)$, etc.) in nuclei. Several years ago, calculations~\cite{Cloet:2005rt,Cloet:2006bq} 
also predicted significant effects for the polarized quark distributions, $\Delta u(x), \Delta d(x)$ and 
the associated structure function, $g_1$. Of particular interest is the fact that nuclear effects 
in the polarized case are predicted to be as large or larger than for the unpolarized structure functions 
with a striking difference in the $x$-dependence of the effect.

Performing measurements of this polarized EMC effect poses certain experimental challenges, however. Both 
polarized beams and targets are required, and experiments using polarized targets are often performed at lower luminosity,
thus requiring longer run times, larger acceptance spectrometers, or both. Additional complications also arise because 
of the presence of unpolarized materials in the target, requiring careful measurements of the ``dilution factor''. 

An intriguing possibility would be to attempt to access the nuclear dependence of polarized quark
distributions via Generalized Parton Distributions (GPDs). In the forward limit, the GPD 
$\tilde{H}^q(x,\xi=0,t=0)$ corresponds to $\Delta q(x)$. GPDs can be probed in hard, exclusive reactions, with 
different processes sensitive to different combinations of GPDs. In the case of the polarized GPD ($\tilde{H}$),
the reaction of interest is exclusive production of pseudoscalar mesons. Experimentally, one would avoid the 
need for a polarized target, however the experimental challenge in this case is to prove soft-hard factorization and
provide clean isolation of the quasi-free reaction.

Despite the experimental difficulties, a measurement of the polarized EMC effect should be a high priority. New
observables such as these are absolutely crucial in differentiating the myriad approaches to describing the 
origins of the EMC effect.

\subsection{Nuclear Dependence of $R=\sigma_L/\sigma_T$}

As discussed earlier, the ratio of cross sections, $\sigma_A/\sigma_D$ can be identified with the ratio of $F_2$ 
structure functions only in the limit $\epsilon=1$ or if the ratio, $R$, of longitudinal to transverse virtual photon
cross sections, $\sigma_L/\sigma_T$ is the same in the nucleus $A$ and deuteron. This has been a topic of much experimental 
investigation (see for example Refs.~\citen{Arneodo:1996ru, Amaudruz:1992wn, Ackerstaff:1999ac, Dasu:1993vk}) since
a non-zero $R_A-R_D$ clouds the interpretation of EMC effect measurements as being sensitive {\emph{only}} to 
modifications of the quark structure functions in nuclei. The common interpretation of the existing experimental 
results is that there is no evidence for a ``significant'' nuclear dependence of $R$. 

It has been demonstrated~\cite{Guzey:2012yk}, however, that the precision of the 
existing data for $R_A-R_D$ is not sufficient to rule out a nuclear dependence that could result in, for example,
the small enhancement of the nuclear ratios in the anti-shadowing region being absent for the ratio of $F_1$ structure
functions and manifesting only in the ratio $F_2^A/F_2^D$. Since $F_1$ contains contributions only from transverse
virtual photons, while $F_2$ contains contributions from both transverse and longitudinal photons, such a
non-zero $R_A-R_D$ implies that anti-shadowing may be a manifestation of a modification of $\sigma_L$ only, i.e., 
not necessarily due to modifications of the quark distributions.

In addition, the only precision data in the large $x$, EMC region comes from SLAC E140, in which an explicit 
Rosenbluth separation was performed. However, in the analysis of that experiment, Coulomb corrections, typically
expected to be small at very high energies, were ignored. In actuality, the relatively low energies needed to fully
span a large lever arm in the virtual photon polarization $\epsilon$ imply that Coulomb corrections must be included.
When the E140 data are re-analyzed with an estimate of the appropriate correction, there are hints that at 
$x=0.5$, the nuclear dependence of $R$ may be non-zero~\cite{Solvignon:2009it}.

The most straightforward way to improve knowledge of $R_A-R_D$ at large $x$ would be to improve upon the E140 Rosenbluth
measurements.  This can be accomplished at Jefferson Lab, taking advantage of the 
high luminosity to  improve the statistical accuracy and extend the $\Delta \epsilon$ lever arm. Of course, 
as in the case of E140, Coulomb corrections can not be ignored and must be handled with care.

An alternate approach to the Rosenbluth technique suggested in Ref.~\citen{Geesaman:1995yd} would be to measure the 
nuclear dependence of the $\cos{\phi}$ moment in semi-inclusive pion production. The $\cos{\phi}$ term in the SIDIS 
cross section is sensitive to the interference cross section, $\sigma_{LT}$ which contains contributions from longitudinal
and transverse photons. While the nuclear dependence of $\sigma_{LT}$ can not be trivially identified with $R_A-R_D$,
it would provide information on whether longitudinal and transverse amplitudes are both modified in the same way in a 
nucleus.

\subsection{Other Measurements}

In addition to the measurements described above, there are even more experiments planned and ongoing that will provide
new information with regards to the EMC effect. The MI$\nu$ERVA experiment~\cite{Drakoulakos:2004gn} at Fermilab will 
provide improved measurements of the nuclear dependence of neutrino cross sections. Modification of the nucleon form 
factor in the nucleus has been another related area of experimentation at Jefferson Lab at 
6~GeV~\cite{Strauch:2002wu,Paolone:2010qc,Malace:2010ft}, with additional measurements planned for the 
12 GeV era~\cite{E12_11_002}.

\section{Summary and Conclusions}

The unambiguous signal presented by the original observation of the EMC effect provided an initial window to 
the manifestation of quark-gluon degrees of freedom in a cold nucleus. The fact that the origin of the nuclear 
modification of quark distributions is still a matter of some controversy thirty years after the original observation
only emphasizes the magnitude of the problem QCD presents.  The path forward in achieving a final and conclusive 
understanding of the EMC effect will clearly involve both experimental and theoretical input.  

It is often presumed that there is not much to be gained from further measurements
of inclusive $A/D$ cross section ratios. However, our analysis of the existing world data for these ratios has 
demonstrated that while certain targets like carbon and iron have been precisely measured over the full range of $x$, there 
are many targets for which additional data is needed. In particular, there are heavier targets that are either well measured
at low $x$ (i.e. Pb and Cu) but not larger $x$, or are well measured at larger $x$ (Ag and Au) but not low $x$.

Beyond filling in gaps in the database of $A/D$ cross section ratios, it is also crucial to pursue new observables.
Several such avenues have been described in the previous section, including measurements of the flavor dependence of the EMC
effect, nuclear modification of polarized quark distributions, etc. Perhaps most crucially, experiments must attempt to further
explore and explain the apparent connection between the modification of quark distributions and nucleon-nucleon
short range correlations in the nucleus. Fortunately, a vigorous experimental program at Jefferson Lab, as well as measurements
at COMPASS and Fermilab will help provide answers in the near future to many of today's outstanding questions. 

These new measurements will also provide new challenges to theories that attempt to describe the EMC effect. Until recently,
there were relatively few observables against which to benchmark particular approaches. The recent observation of the 
EMC-SRC connection, and the subsequent activity attempting to explain its origin only underscores the need for
additional information in the form of new observables.


\section*{Acknowledgements}

The authors would like to acknowledge the many helpful conversations and inputs from 
Alberto Accardi.  This work is supported by the United States Department of Energy's 
Office of Science contract number DE-AC02-06CH11357 under which UChicago Argonne, 
LCC operates Argonne National Laboratory and contract number DE-AC05-06OR23177 
under which Jefferson Science Associates, LCC operates the Thomas Jefferson National 
Accelerator Facility.

\clearpage
\bibliographystyle{ws-ijmpa}
\bibliography{bibliography}
\end{document}